\documentclass[aps,pra,reprint,10pt,a4paper]{revtex4-1}
\usepackage[utf8]{inputenc}
\usepackage[sc,osf]{mathpazo}
\usepackage{amsmath}
\usepackage[T1]{fontenc}
\usepackage{latexsym}
\usepackage{amssymb}
\usepackage[colorlinks=true,citecolor=blue,urlcolor=blue]{hyperref}
\usepackage{color}
\usepackage{graphics,epstopdf}
\usepackage{soul}
\usepackage[demo]{graphicx}
\usepackage{capt-of}
\usepackage{lipsum}
\usepackage{adjustbox}
\usepackage[normalem]{ulem}
\usepackage[table,xcdraw]{xcolor}

\begin{document}
	\title{Constructive Feedback of  Non-Markovianity on Resources in Random Quantum States}

	\author{Rivu Gupta$^{1}$, Shashank Gupta$^{2}$, Shiladitya Mal$^1$, Aditi Sen (De)$^1$}
	
	\affiliation{$^1$ Harish-Chandra Research Institute and HBNI, Chhatnag Road, Jhunsi, Allahabad - 211019, India}
	\affiliation{$^2$ S. N. Bose National Centre for Basic Sciences, Block JD, Sector III, Salt Lake, Kolkata - 700 106, India}

	
	\date{\today}

	\begin{abstract}
		We explore the impact of non-Markovian channels on the quantum correlations of Haar uniformly generated random two-qubit input states with different ranks -- either one of the qubits  (single-sided) or both the qubits independently (double-sided)  are passed through a noisy channel. Under dephasing and depolarizing channels with varying non-Markovian strength, entanglement and quantum discord of the output states collapse and revive with the increase of noise. By both analytical and numerical means,  we find  that in the case of  depolarizing double-sided channel, entanglement and quantum discord for random states shows a higher number of revivals on average than that of the single-sided ones with a fixed non-Markovianity strength, irrespective of the rank of the states -- we call such a counter-intuitive event as  \emph{constructive feedback of non-Markovianity}. On the other hand, the mean value of critical noise at which quantum correlations (QCs) first collapse, decreases with the increase of non-Markovianity, independent of the rank of the random initial states. However, the average noise at which QCs of random states show the first revival decreases with the increase of the strength of non-Markovian noise, thereby indicating the role of non-Markovian channels on the regeneration of QCs even in presence of a high amount of noise.   Moreover,  we observe that the tendency of a state to show regeneration increases with the increase of average QCs of the random input states along with non-Markovianity.
		
	\end{abstract}
	
	
	\maketitle
	
	\section{Introduction} \label{1}
	
	Composite systems in quantum mechanics, described by a tensor product Hilbert space, can show one of the striking nonclassical features called  entanglement \cite{ent}. In particular,  complete information about an entangled pure state can not be determined by the  information of its subsystems. Employing these quantum states, various tasks like teleportation \cite{Bennett1993, Bouwmeester1997}, dense coding \cite{Bennett1992, Mattle1996},  secure key distribution \cite{Ekert1991, Gisin2002, Jennewein2000}, one-way quantum computing  \cite{Rau, Bri} have been designed to achieve higher efficiency than the protocols using unentangled states. Therefore, in the current era, entangled states constitute the basis of cutting edge quantum technologies. On the other hand,  it has also been realized that there are other forms of quantum correlations (QCs) present in quantum states which can  exhibit counterintuitive phenomena, completely inexplicable  by  classical theory. Motivated by classical information theory \cite{ct'91}, quantification of such  correlations leads to a measure called  quantum discord (QD) \cite{QD, discordreviewModi, discordreviewBera}, independent of entanglement and has been identified as a resource in tasks like deterministic quantum computation with single qubit \cite{Datta08}, remote state preparation \cite{rsp, rspHorodecki}, distribution of entanglement \cite{streltsov'11}, quantum locking \cite{locking'11}, identifying quantum phase transition in many-body systems \cite{discordreviewModi, discordreviewBera}. 
	
	Importantly, realizing  all such quantum information processing tasks  in laboratories requires distributing  resource states over space and time.  During this process,  quantum correlations, in general,  get destroyed due to the interactions with the environment, thereby creating obstacles in the successful implementation  of these protocols.  It was found that under local dephasing noise, entanglement disappears suddenly in a finite time,  known as entanglement sudden death (ESD) \cite{suddendeathent} (cf. \cite{FicekTanas}), while other resources like  quantum discord vanish asymptotically \cite{asympdisco, asympdisco1, asympdisco2, discordreviewModi, discordreviewBera}. In contrast to ESD under Markovian noise, the revival of entanglement has also been observed under non-Markovian evolution \cite{RLF17, nonMarkoent}. Moreover, it was shown that there exists a certain class of  states for which  QD remains invariant with the increase of noise even when the system is affected via Markovian or non-Markovian channel-- a phenomenon called freezing of QD  \cite{mazzola'10}. It turns out to be universal for non-dissipative noise under a certain class of states \cite{haikka'13}. In all the previous works, interesting features of the dynamics of entanglement or QD  under noisy environments have been studied for a particular class of initial states \cite{karpat'17}.  
	
	Another interesting avenue of research is to find some pattern in randomly generated states against the intuition of observing random behaviour \cite{Karolbook}.  It was reported that random states can have universal quantum properties like increase of average QCs among randomly generated states with the increase of a number of parties \cite{Karolrandom, Eisertrandom, Winterrandom, Sooryarandom, Ratulrandom, Kendon}. Moreover, random states appear naturally in chaotic systems \cite{haake'10} and have also been employed in disproving a long-standing conjecture in quantum information science regarding additivity of minimal output entropy \cite{hastings'09}.
	
	In the present work, we investigate the effects of local decoherence on QCs of Haar uniformly simulated two-qubit initial states with different ranks and  our aim is to check whether the results obtained for a specific class of states persist even for random states or not. Specifically, when either one of the qubits or both the qubits of random two-qubit states are independently sent through  dephasing and depolarizing non-Markovian channels, we search for generic traits in QCs in the form of entanglement and QD.  Note that such an analysis depends on several factors
	like rank and QC content of the initial states,  properties of channels like strength of the noise and non-Markovianity and it makes the analytical treatment to find the universal patterns  of QCs under decoherence  for random states very difficult. Hence  numerical simulation is a good tool for such investigations although for low ranked states, we can address the problem analytically as well. We find that both the QC measures show revival after the collapse due to the presence of non-Markovianity irrespective of the ranks of  the input quantum states. Interestingly, we show that  if a single qubit is sent through the depolarizing channel, no revival of entanglement is observed while entanglement resurrects after the collapse for a certain amount of non-Markovian noise affecting both the qubits. In a similar spirit, we observe that the mean number of regeneration in  QD  is more when both the qubits are affected by noisy  depolarizing channels compared to the case when a single qubit is sent through it.  We call such a counter-intuitive observation as \emph{constructive effects of non-Markovianity}. This phenomenon is possibly observed due to a competition between the damping and the non-Markovianity in the noisy channel where the former seems to be responsible for the collapse of QCs while the latter is accountable for regeneration.
	Note that such a feature is absent  for output states obtained via dephasing channel. Our analysis also reveals that the double-sided dephasing channel facilitates revival for both entanglement and discord in case of pure states while for states with higher ranks, this is true for both single- and double-sided dephasing channels. 
	
	We also find that for a fixed non-Markovian noise, average value of regeneration in case of entanglement decreases with the increase of the rank of the states, establishing pure states as good resource. 
	Our analysis of mean regeneration and mean noise threshold value for the revival  of random states reveals that non-Markovianity induces regeneration of QCs  even in presence of a high amount of noise in dephasing as well as depolarizing  channels. In contrast, we also notice that the noise-threshold at which entanglement as well as QD initially collapses also decreases with non-Markovianity  and for QD alone, number of states that collapses increases with non-Markovianity which indicates a competition between the strength of non-Markovianity and damping parameters in the channels. This result is in a different spirit than the one which showed that non-Markovianity leads to more number of freezing in QD \cite{karpat'17, discordreviewBera}.  Note that, rank-1 states are maximally robust and QD of significant number of such states does not collapse even at high non-Markovianity. High ranked states also exist whose QDs do not  collapse under both dephasing and depolarizing noise but percentage of such states is less than pure states.
	
	The paper is organised in the following way. In Sec. \ref{sec_prereq}, we recapitulate the Haar uniform generation of two-qubit states of different ranks,  and quantum channels. In Sec. \ref{sec_motiv}, we provide motivation and preliminary observations to proceed further while  Sec. \ref{sec_signi} introduces the significant quantities  required for investigations. In Secs. \ref{sec_deph} and \ref{sec_depol},  results  for random states with  non-Markovian dephasing  and depolarizing channels are respectively presented. 
	Finally, we conclude with a summary of results in Sec. \ref{sec_conclu}.
	
	\section{Prerequisites: Generation of states, and Action of Channels}
	\label{sec_prereq}
	
	Before presenting the results, in this section we will first set the stage by  briefly describing the process of generating Haar uniformly two-qubit density matrices which we  employ here to obtain the input states,  and the action of quantum channels on input states.

	\subsubsection{Generation of Random States}
	\label{subsec_randstate}
	
	We know that, given a basis, quantum states  are  specified by complex coefficients. To generate $5\times 10^4$ states Haar uniformly in state space \cite{Karolbook}, we randomly simulate real  numbers involved in states from a Gaussian distribution with mean $0$ and standard deviation unity, denoted \(G(0,1)\).
	In particular, for a two-qubit pure state,
	\(|\psi \rangle = \sum_{i,j=0,1} (a_{ij} + i b_{ij}) |i \rangle \otimes |j \rangle\),
	$a_{ij}$ and $b_{ij}$ are real numbers chosen \(G(0,1)\). 
	Here $|i\rangle \in \{|0\rangle, |1\rangle\}$ form the computational basis of the first qubit and similarly $|j\rangle$ for the second qubit.  
	Similarly, the set of real numbers,  \(\{a_{ijk}\}\),  \(\{b_{ijk}\}\), \(\{a_{ijkl}\}\), and \(\{b_{ijkl}\}\)  of \(|\psi_3 \rangle = \sum_{i,j,k=0,1} (a_{ijk} + ib_{ijk}) |i,j,k\rangle\), \( |\psi'_3 \rangle = \sum_{i,j=0,1} \sum_{k=0,1,2} (a_{ijk} + ib_{ijk}) |i,j,k\rangle \)  and \(|\psi_4 \rangle = \sum_{i,j,k,l=0,1} (a_{ijkl} + ib_{ijkl}) |i,j,k,l\rangle \), are chosen randomly from \(G(0,1)\) to obtain rank-2 $(R_2)$, rank-3 $(R_3)$ and rank-4 $(R_4)$ two-qubit density matrices  respectively. 
	Notice that  \(|i\rangle, |j\rangle |k\rangle, |l\rangle \in \{|0\rangle, |1\rangle\}\) form the computational basis of qubits \(1\),  \(2\), \(3\) in \(|\psi_3\rangle\), first two qubits of \(|\psi'_3\rangle\) and  all the four qubits of \(|\psi_4\rangle \) while the third party in \(|\psi'_3\rangle\) belongs to the computational basis of qutrit, i.e., \(\{|0\rangle, |1\rangle, |2\rangle\}\). Finally, tracing out a single or two parties leads to the desired two-qubit density matrices with different ranks. 
	%
	%

	\subsection{Non-Markovian quantum channels}

	
	Let us consider two paradigmatic  noisy channels, namely dephasing and depolarizing channels whose
	Kraus operators with the strength of non-Markovianity, $\alpha$ \cite{dephNM, depolarNM}, are respectively given by 
	\begin{eqnarray}
		\label{Krausnoise}
		&& K_{I}^{dph} =\sqrt{[1-\alpha p](1-p)} I, K_{z}^{dph} = \sqrt{[1+\alpha(1-p)]p}\sigma_{z}, \nonumber \\
		&& K_{I}^{dp} =\sqrt{[1-3\alpha p](1-p)} I,  K_{i}^{dp} = \sqrt{\frac{[1+3\alpha(1-p)]p}{3}}\sigma_{i}. \nonumber\\
	\end{eqnarray}
	Here $0\leq \alpha \leq1$,  and $\sigma_i,\, i = x,y,z$ are the Pauli matrices. In case of dephasing channel, $0\leq p\leq 0.5$ while in depolarizing case,  $0\leq p\leq 1$. Note that $\alpha = 0$ represents the Markovian case \cite{Preskillnotes, NielsenChuang}.
	Our aim is to study the trends of QCs of the output states when randomly generated input states having different ranks are subjected to the non-Markovian dephasing and depolarizing channels. Towards this analysis, we consider following two scenarios:
	\begin{enumerate}
		\item[] \emph{Situation 1.}  Noise acts on one of the sites of the two-qubit state, which we refer to as the \emph{single- sided channel}.
		Let $\rho_{0}$ be the initial state. After subjecting it to the single-sided channel, the resulting state can be represented as
		\begin{equation}
			\rho_{0} \rightarrow \rho_{f}(p) = \sum_{i} (K_{i}^{n} \otimes I) \rho_{0} (K_{i}^{n} \otimes I)^{\dagger} ,
		\end{equation}
		where $K_i^{n}$s are Kraus operators corresponding to either dephasing or depolarizing channel.
		\item[] \emph{Situation 2.} When both the parties are sent through two local channels, we call it as a \emph{doubled-sided channel}. The output state in this case  reads as 
		\begin{equation}
			\rho_{0} \rightarrow \rho_{f}(p) = \sum_{i, j} (K_{i}^{n} \otimes K_{j}^{n}) \rho_{0} (K_{i}^{n} \otimes K_{j}^{n})^{\dagger}.
		\end{equation}
	\end{enumerate}

	\section{Motivation: Preliminary observations}
	\label{sec_motiv}
	
	Before moving to the study of quantum correlations in random states under Markovian and non-Markovian channels, let us discuss some of the counter-intuitive results  known in literature as well as some specific exemplary cases. These results motivate us to look for generic features in QC for random states under decoherence.

	
	First of all, it was realized in different studies that entanglement is, in general, fragile in a noisy scenario, i.e. entanglement decays with the increase of noise. Moreover, it was  demonstrated that for a certain class of  states,  entanglement suddenly vanishes at a fixed noise parameter, referred as sudden death of entanglement \cite{suddendeathent}. At the same time, in presence of non-Markovian  and common  (Markovian) noise models, it was shown that entanglement collapses as well as revives with the increase of noise strength \cite{RLF17, FicekTanas, nonMarkoent}. 
	Let us now prove that such a revival of entanglement cannot be seen in case of Markovian dephasing channel when the input state, \(\rho_{AB}^{BD}\), shared between Alice and Bob,  is the class of Bell diagonal (BD) states, given by 
	\(\rho_{AB}^{BD} = \frac{1}{4} (I_4 + \sum_{i=x, y, z} C_{ii} \sigma_i \otimes \sigma_i)\),
	where \(C_{ii} = \mbox{tr}( \rho_{AB}^{BD} \sigma_i \otimes \sigma_i)\) are the classical correlators. It reduces to the singlet when \(C_{xx}  = - C_{yy} = 1\) and \(C_{zz} =1\).

	\textbf{Proposition 1.}: \emph{When the dephasing Markovian channel acts on a single qubit or on both the qubits locally of a two-qubit Bell diagonal state with \(C_{xx} =1,\,\,C_{yy}=-C_{zz}\), the entanglement of the resulting  state always vanishes, thereby showing no revival of entanglement with noise.  } \\
	
	\emph{Proof.} Suppose first that  dephasing noise (Eq. (\ref{Krausnoise}) with \(\alpha =0\)) acts on a single qubit, i.e., at Alice's side and  the resulting state
	under the action of dephasing noise
	is  given by
	\begin{equation}
		\rho_{AB}'=\left(
		\begin{array}{cccc}
			\frac{1+C_{zz}}{4} & 0 & 0 & A \\
			0 & \frac{1-C_{zz}}{4} & B & 0 \\
			0 & B& \frac{1-C_{zz}}{4} & 0 \\
			A & 0 & 0 & \frac{1+C_{zz}}{4},
		\end{array}
		\right)
	\end{equation}
	where $A = -\frac{1}{4} (C_{xx}-C_{yy}) (-1+2 p)$ and \(B = -\frac{1}{4} (C_{xx}+C_{yy}) (-1+2 p) \).
	The eigenvalues of partially transposed state $\rho_{AB}^{'T_{A}}$ for $C_{xx}=1$ and $C_{yy}=-C_{zz}$ are
	\(\lambda_{A1}=\frac{1}{2} (1+(-1+C_{zz}) p), \, 
	\lambda_{A2}=\frac{1}{2} (C_{zz} (-1+p)+p), 
	\lambda_{A3}=\frac{1}{2} (1-(1+C_{zz}) p) ,
	\lambda_{A4}=\frac{1}{2} (C_{zz}+p-C_{zz} p)\). 
	The two-qubit state is entangled only when one of the eigenvalues is negative. Such a possibility occurs in 
	following two cases: 
	1. For $-1 \leq C_{zz} < 0,\, \lambda_{A4}$ turns out to be the minimum among  four eigenvalues. $\lambda_{A4}$ goes to zero from negative value when $p\to \frac{C_{zz}}{-1+C_{zz}}$ and beyond this value, all eigenvalues remain positive. It, therefore, implies that  p has just a single value upto which entanglement is nonvanishing for any values of  $C_{zz}$, and hence no entanglement is present in the output state after the first collapse.
	2. When $0 < C_{zz} \leq 1, \, \lambda_{A2}$ can only be negative and it goes to zero when $p\to \frac{C_{zz}}{1+C_{zz}}$ which again leads to the fact that entanglement survives only when $p< \frac{C_{zz}}{1+C_{zz}}$ and the rest of the range of $p$,  the state is unentangled, thereby showing no revival of entanglement.
	
	Let us now move to the case when  the local dephasing noise acts on both the sides of  the initial state, leading to the resulting state which is in the same form as single-sided one except  the off-diagonal element now become \(X'= X(1-2 p),\, X= A, B\). 
	The eigenvalues of partially transposed state, $\rho_{AB}^{''T_{A}}$, in this case, reads as
	\(\lambda_{AB1,2}= (1-p) p \pm C_{zz} \left(\frac{1}{2}-p+p^2\right), \lambda_{AB3,4}=\frac{1}{2} -(1\pm C_{zz}) p + (1 \pm C_{zz}) p^2\).  Like single-side case, similarly two situations arise and can be shown that there is no revival of entanglement. 
	\hfill $\blacksquare$
	
	\begin{figure}[!ht]
		\resizebox{7.5cm}{5.5cm}{\includegraphics{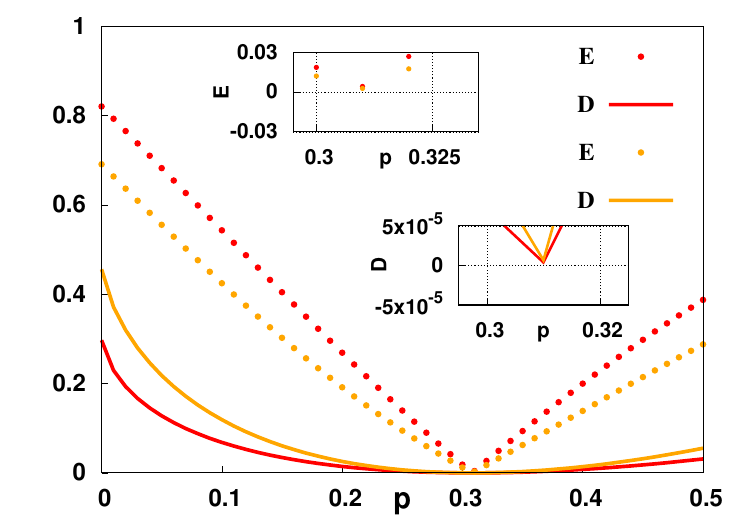}}
		\caption{(Color online.)  Behavior of QCs of two output states (vertical axis)  against noise, $p$ (horizontal axis). The output states are obtained when two random pure states \cite{randompure} are sent through a single-sided dephasing channel.  Here we set non-Markovianity parameter,  $\alpha = 0.9$. Dotted lines represent LN while solid  ones are for  QD. Insets show that both LN and QD  do not collapse for these states.  The horizontal axis is dimensionless while LN and  QD are  in ebits and bits respectively. }
		\label{motiv}
	\end{figure}
	
	\emph{Remark 1.} Although we have given the proof for dephasing channel, similar results can be obtained for Markovian depolarizing channel with the Bell diagonal states as input. 
	
	\emph{Remark 2.} When (Markovian) dephasing and depolarizing single- and double-sided channels act on randomly generated two-qubit states with different ranks, we find that the above Proposition for entanglement remains valid. 
	
	\emph{Remark 3.} The Proposition also holds for quantum discord, only in case of the (Markovian) dephasing channel.

	\emph{Remark 4. }  Interestingly, we  observe that LN as well as QD for randomly generated pure states after  sending through single-sided non-Markovian dephasing channel  do not vanish (upto numerical accuracy $10^{-6}$), (see illustration of two random pure states in Fig. \ref{motiv}). We will address this issue in details in succeeding section. 
	
	\begin{figure}[!ht]
		\resizebox{8cm}{5.5cm}{\includegraphics{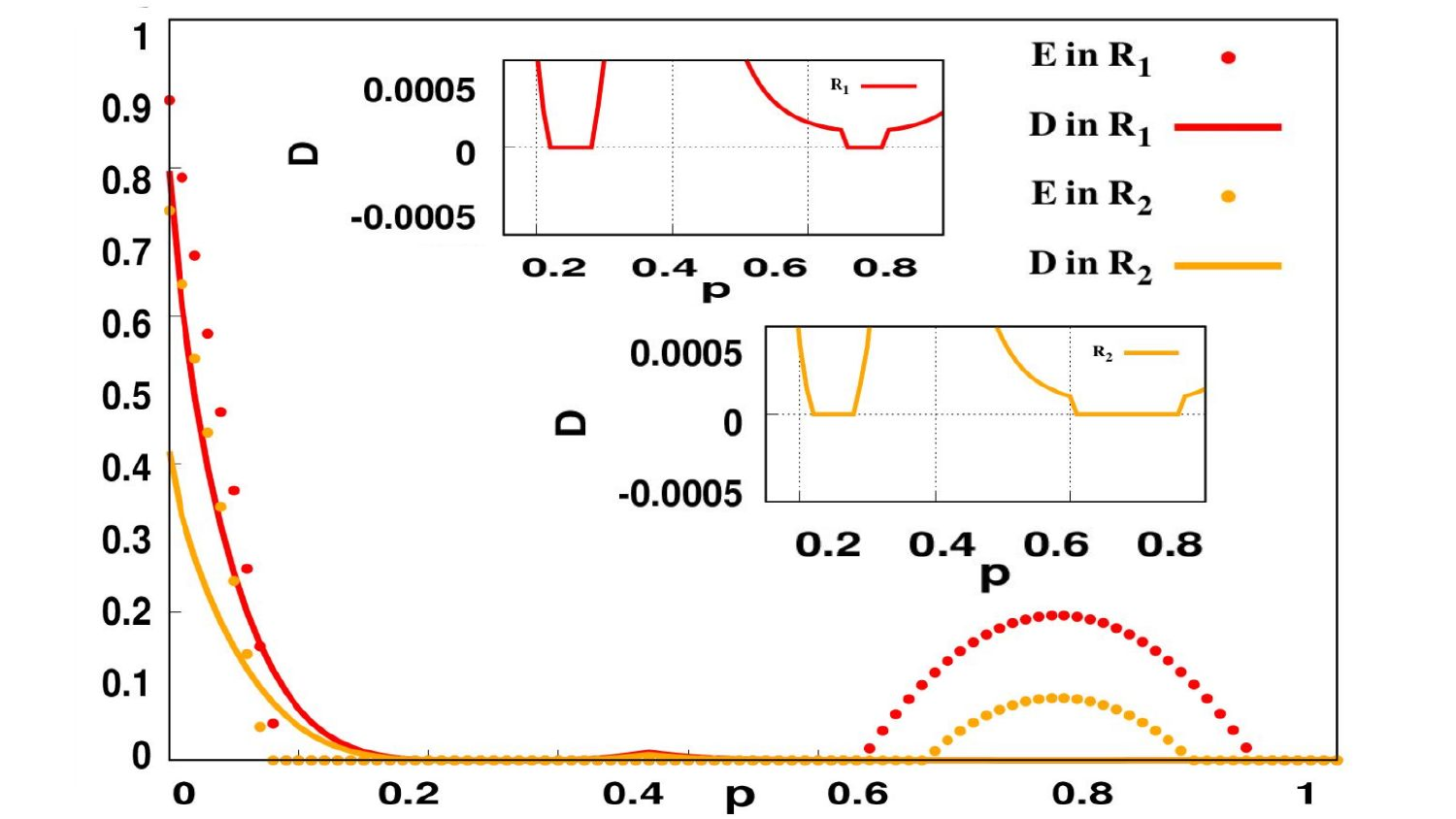}}
		\caption{ (Color online.) QCs of the resulting states after passing through the  double-sided depolarizing channels  vs.  $p$. Dotted dark (red) and light (orange) lines are for LN of rank-1 \cite{randompuremix} and rank-2 states respectively  while solid dark (red) and light (orange) are for QD of the initial \(R_1\) and \(R_2\) states. 
			Here $\alpha$ = 0.9. QCs show  revivals after an initial collapse and again collapse with the increase of the damping parameter, $p$. All  other specifications are same as Fig. \ref{motiv}. The two collapse for QD has been better shown in the insets.
		}
		\label{motivation}
	\end{figure}

	Contrary to the  observation in Fig. \ref{motiv},  we note that   when non-Markovian depolarizing channel acts on both the qubits of  low rank randomly simulated states (pure as well as rank-2 states), LN and QD show two collapses, i.e., they  collapse, revive and further collapse (see  Fig. \ref{motivation}). It implies that several collapses and revivals can occur under non-Markovian channels. We will carefully observe in the succeeding section whether such exotic behavior of QC can  have some connection with initial amount of QC present in the state and other characteristic of the state or the channel. 
	

	

	\section{Significant Quantities introduced for investigations}
	\label{sec_signi}
	
	As it was known and also seen from the preceding section, QC of the output state can, in general,  show collapse as well as revival with the variation of noise parameter. Towards finding the universal feature of QCs in the resulting state from random input states, we define here a  few physical quantities which will help us to perform the analysis.
	
	As shown in Fig. \ref{motivation}, collapse and regeneration of entanglement as well as QD can occur more than once with noise, especially with Non-Markovian channels.  As it will be clear from the analysis in the succeeding sections, there is a competition between noise and non-Markovianity which affects the behavior of  QC measures. 
	Towards establishing a connection between 
	non-Markovianity on  states and the content of QC of the input state which is necessary for  regeneration,  we introduce a quantity which we call normalized regeneration.
	
	\emph{Normalized regeneration.} For a fixed value of non-Markovianity, $\alpha$, and a fixed rank of the input state, normalised regeneration is defined as the ratio between the number of  regeneration shown by a state having a fixed amount of  QC i.e., $a \leq \mathcal{Q} \leq b$, with $\mathcal{Q}$ being the measure of QC and the total number of generated  states, within that range of $\mathcal{Q}$ that shows collapses. Mathematically,
	\begin{equation}
		\label{eq_normreg}
		R_g^N = \frac{\mbox{Number of regeneration with inputs in} \,\mathcal{Q} \in (a, b) }{\mathcal{N}_{\mathcal{Q} \in (a, b)} },
	\end{equation}
	where  \( \mathcal{N}_{\mathcal{Q} \in (a, b)}\) is the total number of output states obtained from the input states having \( \mathcal{Q} \in (a, b) \) that vanishes in this range. \(\mathcal{Q}\) is either LN or QD. Note that there are significant number of two-qubit states of all ranks under both kind of noisy channels whose QD does not collapse, so, there is no question of regeneration arises for such states. Entanglement collapses in every case except for rank-1 states under the single-sided dephasing channel for all values of non-Markovian parameter $\alpha$. \\
	
	Notice that apart from capturing the role of non-Markovianity, the normalized regeneration is also introduced to capture the effect of initial correlation content of the states which have a tendency to show revival. We will illustrate in the following section that the number of random states generated with high value of non-classical correlation decreases with increasing rank. Only with this result, it seems that on average rank-1 states have more tendency to show revival in comparison to rank-3 states although  it is not the complete picture, as will be evident from our results throughout the discussion. Hence, the normalized regeneration is introduced to study the number of revivals in a range of initial correlation divided by the number of states that are generated within that range of initial QC and thus we are able to remove the effect of rank from the study of regeneration.\\

	\emph{Mean Regeneration. } Based on the normalized regeneration and a fixed QC, \(\mathcal{Q}\),  mean regeneration denoted by  \(\overline{R}_g^{\mathcal{Q}}\)  for a given \(\alpha\) and for   a fixed rank of the input states is defined as
	\begin{equation}
		\label{eq_meanreg}
		\overline{R}_g^{\mathcal{Q}} = \frac{\sum_i R_{g_{i}}^N}{\mathcal{N}},
	\end{equation}
	where \(R_{g_{i}}^N\) is the normalized regeneration observed for the input state possessing QC between \(a_i\) and \(b_i\), summation is over all such quantities in the entire range of $\mathcal{Q}$, i.e., $0\leq \mathcal{Q}\leq 1$, and \(\mathcal{N}\) is the total number of Haar uniformly generated state for a given rank that collapse. In this paper, we always take \(b_i-a_i =0.1\) for all \(i\). This quantity has been introduced to study the effect of  rank, initial correlation of states showing revival and  non-Markovianity on average for a fixed rank and non-Markovianity strength. 
	%
	
	\emph{Mean critical noise for collapse. }
	We are interested in the strength of noise at which quantum correlations vanish for the first time on average for random states, i.e. mean noise threshold when the first collapses of \(\mathcal{Q}\) occur. The critical value of noise quantifies the detrimental effects of noise on system although  it can have some connection with the rank of a state and the strength of non-Markovianity. Towards answering these questions, for a given QC measure, $\mathcal{Q}$, we define a quantity called mean critical value of noise for collapse, denoted by \(\overline{p}_{c}^{\mathcal{Q}}\), as
	\begin{equation}
		\overline{p}_{c}^{\mathcal{Q}} = \frac{\sum_{i=states} p_{c}^i}{\mbox{Total number of generated states that collapsed}},
	\end{equation}
	where \(p_c^i\) denotes  the threshold value of noise at which a QC measure of a given state first collapses and the summation is over all such generated states for a fixed rank showing collapse for $\mathcal{Q}$. For a given non-Markovianity and for a fixed rank, \(\overline{p}_{c}^Q\) determines a \emph{universal} robustness of $\mathcal{Q}$ which the random states possess against a specific noise.
	As we have argued in Sec. \ref{sec_motiv}, QCs of pure states never collapse for single-sided dephasing channel and so \(\overline{p}_{c}^{\mathcal{Q}}\) does not exist for randomly generated pure states for all values of \(\alpha\) in dephasing channel upto the numerical precision. \\

	\emph{Mean critical noise for regeneration.} There can be some inherent characteristics of quantum input states as well  as quantum channels which induce QCs of the state to revive  after collapse.  We are interested to obtain a pattern of first revival or regeneration of QC among random states.  Since it captures the advantage of non-Markovianity on states, it is a kind of complementary measure than the mean critical noise for collapse. For a given QC, \(\mathcal{Q}\), one can also expect an association of \(\overline{p}_{c}^{\mathcal{Q}}\) with the first revival. To seek such a relation, for  a fixed \(\mathcal{Q}\), we have
	\begin{equation}
		\overline{p}_{reg}^{\mathcal{Q}} = \frac{\sum_{i=states} p_{reg}^i}{\mbox{Total no. of simuated states  showing revival}},
	\end{equation}
	where $p_{reg}^i$ denotes the inherent noise of the channel, at which  quantum correlation becomes nonvanishing after the first collapse and summation is over all states which show regeneration. \\

	\emph{Mean initial QC.} Let us finally identify a quantity based on a QC measure which can answer whether there is any lower bound on the content of  QCs in the initial states which can show revival after collapse in presence of non-Markovian channel. 
	Note that it has a meaningful interpretation if for a fixed rank, QC of all the states generated  shows revival  with a fixed non-Markovianity strength.  An average amount of  QC of the input states responsible for regeneration can be defined as follows: For a given \(\mathcal{Q}\), we have
	\begin{equation}
		\overline{\mathcal{Q}}_{in} = \frac{\sum \mbox{QC of states showing regeneration}}{\mathcal{N}}.
	\end{equation}
	Here the summation is over all Haar uniformly generated states which show regeneration. It has to be computed for a fixed $\alpha$ and for a fixed rank of random states. \\
	
	\textit{Remark. } The intuition behind introducing
	these quantities is to separate out the constructive and destructive feedback of noise on states. For example, the mean and  the normalized regeneration as well as  the mean critical noise for regeneration capture the constructive effect of non-Markovianity while the mean critical noise for collapse captures the destructive effect of noisy channels. All these quantities manifest the competition between the damping effect and non-Markovianity strength of the noisy channels.

	\section{Effects of non-Markovian Dephasing Channel on Quantum correlations of Random States}
	\label{sec_deph}
	
\begin{figure}[!ht]
		\resizebox{8.5cm}{6cm}{\includegraphics{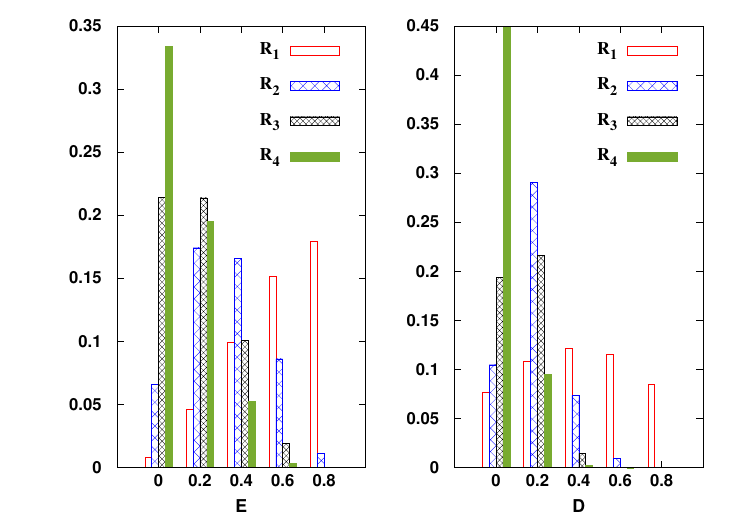}}
		\caption{\footnotesize (Color online.) Normalized frequency distribution of random two-qubit states (vertical axis) against LN (horizontal axis in left) and QD (\(x\)-axis in right). We Haar uniformly generate these states with different ranks as mentioned in Sec. \ref{subsec_randstate}. These states are used as  the initial states before sending  through the noisy  channels in  succeeding sections. The hollow columns represent rank-1, the dashed columns represent rank-2, the checkered columns represent rank-3 and the solid columns represent rank-4 states. All the axes are dimensionless.   }
		\label{state1}
	\end{figure}

	In this section, the non-Markovian dephasing channel acts on a single qubit or both the qubits of randomly generated rank-1 to rank-4 two-qubit states.   The role  of non-Markovianity on QCs of states having different ranks will be estimated by using the previously introduced quantities. As we will show, we can make a general inference about the behavior of QCs in presence of non-Markovian noise. Before stating the results,  let us first ask the following question:  for a given rank, what is the frequency distribution of LN and QD for random states? The pattern of normalised frequency distribution \cite{normdistri}  for  LN and QD of randomly generated two-qubit states with different ranks is depicted in Fig. \ref{state1}. Note that these states are used as the input states before the action of decohering channels.    
	
	
	We see that  average QCs in the Haar uniformly simulated random states decreases with the increase of the rank which is in a good agreement with the  previously known results in Refs. \cite{Karolrandom, Eisertrandom, Winterrandom, Sooryarandom, Ratulrandom}. It was shown that  average multipartite QCs in random pure states increases with the increase of number of parties, thereby showing that almost all multiqubit pure states are highly quantum correlated,  independent of the choice of the quantum correlation measure.  Monogamy of quantum correlations \cite{monogamy} along with the results on random multipartite states implies  that average QCs in two-qubit random density matrices should decrease with the increase of the rank of states, thereby confirming the results in Fig. \ref{state1}.
	
	Let us now analytically show some of the observations obtained numerically for entanglement and discord after action of non-Markovian dephasing channel. Since any analytical calculations involve all the parameters in states as well as  channels, obtaining QCs for the resulting states become extremely difficult. We will show below that for low ranked states and for a restrictive scenario, some analytical treatments are possible. In particular, 
	instead of performing optimizations over all single-party measurements in the computation of QD, 
	we restrict ourselves to the set of projective measurements,  namely, \(\{|0\rangle, |1\rangle, \frac{1}{\sqrt{2}} (|0 \rangle \pm |1\rangle,  \frac{1}{\sqrt{2}} (|0 \rangle \pm i |1\rangle \}\) and calculate QD which we refer as restricted QD. In the next sections, we will show that the results obtained via restricted QD match with the actual trends of QD obtained numerically.

	\textbf{Proposition 2.} \textit{Entanglement and restricted quantum discord of a pure two-qubit state undergo collapse as well as regeneration, when subjected to the double-sided dephasing channel, for a significantly high value of the non-Markovianity parameter.} \\
	
	\textit{Proof.} Any two-qubit pure state may be written as \cite{NielsenChuang}
	\(|\psi_{AB}\rangle = \cos \frac{\theta}{2} |00\rangle + \sin \frac{\theta}{2}|11 \rangle\),
	where $|0\rangle$ and $|1\rangle$ represent  orthonormal bases. After passing through the non-Markovian dephasing channel, the final state $\rho_f = \sum_{i,j} (K_i^{dph} \otimes K_j^{dph}) \rho (K_i^{dph} \otimes K_j^{dph})^\dagger$ and its entanglement as well as QD are functions of $\theta, \alpha$ and $p$. 
	Solving for the zeros of the negativity in terms of the noise parameter, we find its expression as
	\(p_0 = \frac{(1 + \alpha -  \sqrt{1 + \alpha^2})}{\alpha}f(\theta,\alpha)\), which ensures the collapse of entanglement. 
	To show  its regeneration, we calculate the negativity at $p = p_0 \pm h$  as $h \to 0$. If the negativity can be shown to be positive at both these limits for a fixed values of \(p\) and \(\theta\), it means that entanglement is nonvanishing before collapse and  it again regenerates.
	
	For $p = p_0 \pm h$,  we find that the negativity varies as $2h^2 \sin\theta (1 + \alpha^2)$ upto $\mathcal{O}(h^2)$. This is sufficient to show the regeneration of entanglement, since  the entanglement remains positive both to the left and to the right of $p_0$. Furthermore, since the negativity is proportional to $\alpha^2$ for $p = p_0 + h$, it implies that with the increase of  the non-Markovianity, the regenerated value of entanglement increases, thereby highlighting the beneficial effects of non-Markovianity on regeneration. \\
	
	We perform the similar analysis in  case of  restricted QD.
	Although the compact form of QD is cumbersome, for a fixed values of \(\theta\) and a high value of \(\alpha\), the regeneration can be confirmed. For example, 
	when $\cos\frac{\theta}{2} = 0.4$, 
	$\alpha = 0.9$,   $p_c^D = 0.26$ while $p_{reg}^D = 0.434$.   \hfill $\blacksquare$ \\
	
	With detailed numerical analysis, we will show in the following sections  that both $\overline{p}_c^D$ and \(\overline{p}_{reg}^D\) decreases with an increase in the non-Markovianity parameter.
	
	\textbf{Proposition 3.}\textit{ Restricted discord of a two-qubit rank-2 state collapses and regenerates, both for the double-sided and single-sided dephasing channels.}\\
	
\begin{figure}[!ht]
		\resizebox{8.5cm}{5.5cm}{\includegraphics{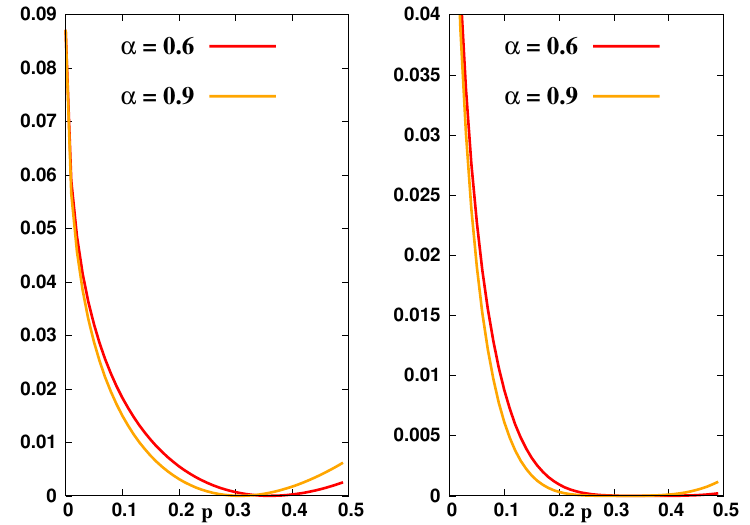}}
		\caption{(Color online.) Behavior of restricted QD after the actions of single-sided (left) and double-sided (right) dephasing channels for $\alpha = 0.6$ (dark) and $\alpha = 0.9$ (light). The state parameters are $\cos \theta_1 = 0.2$, $\cos \theta_2 = 0.4$ and $p_1 = 0.6$.  Both the axes are dimensionless. }
		\label{theorem}
	\end{figure}

	\textit{Proof.} Any rank-2 two-qubit state can be written as
	\begin{equation}
		\rho_{AB}^2 = p_1 |\psi_1 \rangle \langle \psi_1| + (1 - p_1)|\psi_2 \rangle \langle \psi_2|,
		\label{r2state}
	\end{equation}
	where $0 \leq p_1 \leq 1$, $|\psi_1\rangle = |0 \eta_1\rangle + |1 \eta_2 \rangle$ and $|\psi_2\rangle = |0 \eta_1^\perp\rangle + |1 \eta_2^\perp \rangle$ are two mutually orthogonal states \cite{rank2} with $|\eta_1 \rangle = \cos\frac{\theta_1}{2}|0\rangle + \sin\frac{\theta_1}{2} |1\rangle$ and $|\eta_2 \rangle = \cos\frac{\theta_2}{2}|0\rangle + \sin\frac{\theta_2}{2} |1\rangle$, and  $|\eta_i^\perp \rangle$ being states orthogonal to $|\eta_i \rangle$ \((i=1,2)\). QD is again functions of state parameters,\(\theta_i\)s,  \(p_1\),   and parameters involved in the channel which makes the expression cumbersome. After careful analysis, one can confirm the regeneration as shown in Fig. \ref{theorem})   
	for the exemplary values of  state and channel parameters.
	
	
	\textit{Remark.} When the non-Markovian dephasing channel  acts on rank-2 two-qubit states, the entanglement also resurrects after collapse as we will also see in the next subsection.

	\subsection{Constructive outcome of non-Markovian noise on random states}
	
	Let us now analyse the similar and the complementary patterns that entanglement and quantum discord of random states show under decoherence.  
	We start with  entanglement when a fixed amount of non-Markovianty is present in the channel and  the input states are chosen from  a fixed rank. 
	The observations can be divided in two categories, one for single- and another for double-sided channels. We first address the issue of regeneration of entanglement, with the increase of noise and non-Markovianity. 
	
	\begin{enumerate}
		\item \emph{Single-sided channel with pure states.} When a single qubit of random pure states passes through a Markovian channel (i.e., \(\alpha =0\)), LN  never vanishes  with the increase of $p$ except at the point $p=0.5$. With the increase of non-Markovianity, we see that the trend of LN remains almost same. Specifically, we find that for all values of \(\alpha\), LN of all states decreases  to a minimum value with increasing $p$,  then starts increasing with $p$. Note, however, that the minimum of LN  decreases with the increase of $\alpha$.
		
		\begin{figure}[h]
			\resizebox{9.1cm}{7cm}{\includegraphics{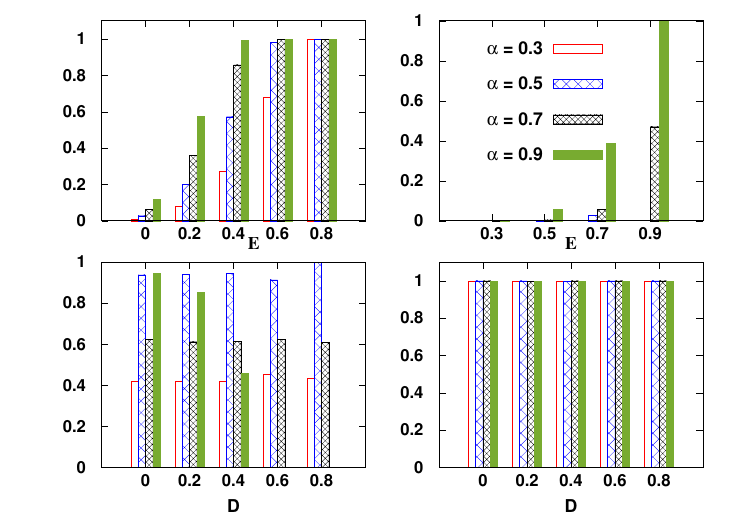}}
			\caption{ (Color online.) Distributions of normalised regeneration, $R_g^N$, in Eq. (\ref{eq_normreg}),  (ordinate) with 
				respect to initial QCs (abscissa). The input states are random rank-2 states which are sent through the dephasing channels for different non-Markovian parameters. Upper panels are for LN while lower panels are for QD. Plots in the left and the right columns are for single- and double-sided channels. The hollow columns represent $\alpha = 0.3$, the dashed columns represent $\alpha = 0.5$, the checkered columns are for $\alpha = 0.7$ while the solid columns represent $\alpha = 0.9$. Quantities in ordinate is dimensionless while \(E\)  and \(D\) are in ebits and bits. Note that no state is generated with QD $\geq$ 0.9. 
			}
			\label{NormR}
		\end{figure}

		\begin{table}[]
			\resizebox{0.5\textwidth}{!}
			{\begin{minipage}{0.7\textwidth}
					\caption{$\overline{R}_g^E$ (Dephasing channel)--Mean regeneration of entanglement when two-qubit states are passed through single- and double-sided dephasing channels is tabulated for different non-Markovianity parameters. It increases with the increase of non-Markovianity although for the double-sided channels, the  effects of noise is so destructive that most of the states do not show revival.  } 
					\label{tabmeanreg}
					\centering 
					\begin{tabular}{|l|l|l|l|l|l|l|l|l|}
						\hline
						\multicolumn{8}{|c|}{$\overline{R}_{g}^E$}                                                                                                                                                                                                                                       \\ \hline
						\multicolumn{1}{|c|}{}         & \multicolumn{1}{c|}{Rank 1}                               & \multicolumn{2}{c|}{Rank 2}                               & \multicolumn{2}{c|}{Rank 3}                               & \multicolumn{2}{c|}{Rank 4}                               \\ \hline
						\multicolumn{1}{|c|}{$\alpha$} & \multicolumn{1}{c|}{double} & \multicolumn{1}{c|}{single} & \multicolumn{1}{c|}{double} & \multicolumn{1}{c|}{single} & \multicolumn{1}{c|}{double} & \multicolumn{1}{c|}{single} & \multicolumn{1}{c|}{double} \\ \hline
						&                             &                             &                             &                             &                             &                             &                             \\ \hline
						0                                                & 0                     & 0                     & 0                      & 0                     & 0                     & 0                     & 0                     \\ \hline
						0.2                                                & 0                     & 0.159                     & 0                      & 0.011                     & 0                     & 0.0004                     & 0                     \\ \hline
						0.3                                                & 0.015                     & 0.26                     & 0                      & 0.032                     & 0                     & 0.003                     & 0                     \\ \hline
						0.5                                                & 0.046                     & 0.458                     & 0.001                     & 0.115                     & 0                     & 0.028                    & 0                     \\ \hline
						0.6                                                & 0.088                     & 0.545                     & 0.002                     & 0.18                     & 0                     & 0.053                     & 0                     \\ \hline
						0.7                                             & 0.166                     & 0.615                     & 0.007                    & 0.241                     & 0                     & 0.089                      & 0                     \\ \hline
						0.8                                                & 0.305                     & 0.682                     & 0.02                      & 0.312                     & 0.002                     & 0.136                     & 0                     \\ \hline
						0.9                                              & 0.545                     & 0.738                     & 0.051                     & 0.384                     & 0.005                     & 0.192                     & 0.0005                     \\ \hline
					\end{tabular}
			\end{minipage}}
		\end{table}
		
		\begin{figure}[!ht]
			\resizebox{9cm}{6cm}{\includegraphics{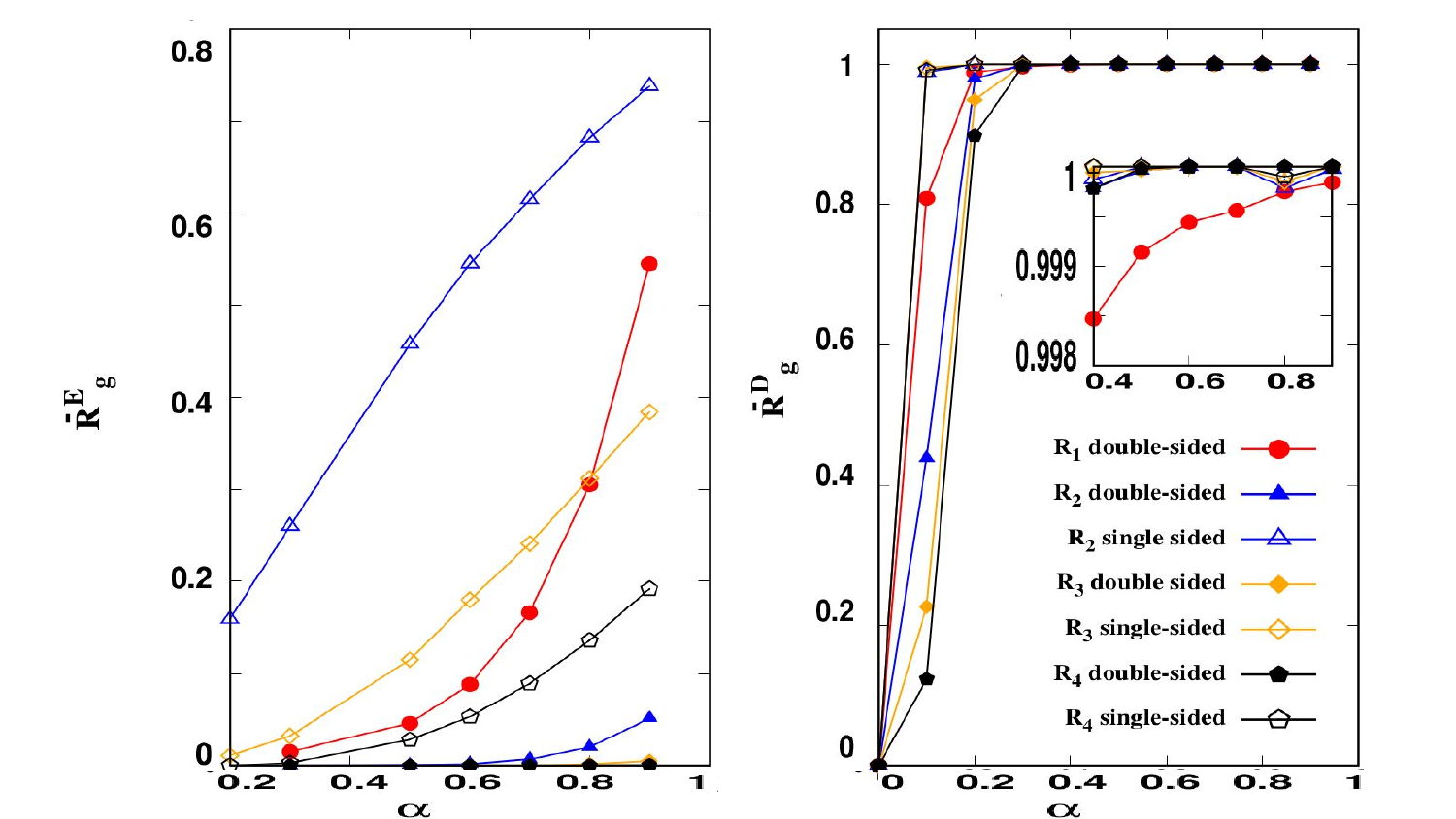}}
			\caption{(Color online. ) Mean regeneration,  $ (\overline{R}_g^Q)$, vs. non-Markovianity,  \(\alpha\).  The hollow and the solid symbols represent single- and double-sided dephasing channels respectively while the left and the right ones are respectively for LN and QD. Ranks-1, 2, 3 and 4 are shown with circles, triangles, diamonds and pentagons respectively.  All quantities plotted are dimensionless. Inset in the right hand figure shows the slight variation of mean regeneration with non-Markovianity for discord.
			}
			\label{figmeanregent}
		\end{figure}
		
		\item \emph{Input states with high rank via single-sided channel.} LN of random states with rank-2, rank-3 and rank-4 show qualitatively similar behavior. For low values of $\alpha$, LN of almost all states does not revive after collapse while entanglement of large fraction of generated states show revival with moderate presence of non-Markovian noise. Note, however, that the regeneration of LN also depends on the rank  and  the entanglement content of the input states. Specifically,  certain value of non-Markovianity in the channel together with the entanglement of the initial states leads to the revival of entanglement as shown in Fig. \ref{NormR} for rank-2 states. For example, we observe that in case of rank-2 states, if LN of the initial state is higher than \(0.8\), the resulting states with a moderate value of   non-Markovian strength  like  \(0.3\) always revive after collapse, thereby showing normalized regeneration to be unity as shown in the upper left panel in Fig. \ref{NormR}.  This observation also matches with the pattern of mean regeneration given in Table. \ref{tabmeanreg} and Fig. \ref{figmeanregent}. In particular, we find that \(\overline{R}_g^E\) increases with the increase of \(\alpha\) monotonically for a fixed rank of the input state, implying that percentage of randomly generated states showing regeneration after collapse increases with the increase of non-Markovianity.   For example, we find that with \(\alpha =0.9\), LN of only \(19.2\%\) $R_4$ states can again revive while it is \(73.8\%\) and \(38.4\%\) respectively for $R_2$ and $R_3$ states. These percentages have to be considered along with the frequency distribution of entanglement for input states with ranks in Fig. 3, i.e., regeneration of entanglement also depends on the entanglement content of the initial states. 
		These results indicate a complex relation of collapse and revival of entanglement with the critical values of non-Markovianity and noise strength as well as the initial value of entanglement of the input states. We will shed light on these issues when we consider the mean critical noise for collapse and regeneration as well as mean entanglement content of random states.

		
		\item \emph{Action of double-sided dephasing channel on entanglement. } When both the parties are effected by local non-Markovian noise, collapse followed by a revival of entanglement can be seen in random  states with all the ranks for moderate value of $\alpha$. In general, normalized regeneration starts increasing with the increase of non-Markovian strength although the input state must possess moderate to high amount of entanglement. If we fix the rank of states and non-Markovian parameter, we observe  that double-sided channel has detrimental effects on LN compared to a single-sided channel (comparing left and right panels in Fig.  \ref{NormR} ) -- we can refer this as \emph{destructive effect} of noise on entanglement. Moreover, we find that for a fixed \(\alpha\) and rank of the states, mean regeneration in this case  is also very low compared to a single-sided channel. For a given \(\alpha\),  we notice that pure input states have maximum \(\overline{R}_g^E\) as depicted in Fig. \ref{figmeanregent}.
		
	\end{enumerate}

	Let us now move to the behavior of quantum discord in above situations and examine whether similar picture reported for LN emerges for QD or not. 
	\begin{enumerate}
		
		\item \emph{QD under the action of single-sided channel. } In case of pure  random states as inputs, we have already shown that entanglement of the resulting states in almost all cases does not collapse for any values of non-Markovian strength of the dephasing  channel under single-sided action.  From the definition of QD, we know that QD is positive for all entangled states \cite{QD} and hence it also does not vanish. From our numerical simulations, we find that although the trend of QD  is similar to that of LN, the minimum value attained by it is one order of magnitude lower than LN. \\
		
		In case of random states with $R_2$, $R_3$, and $R_4$, we observe a behaviour similar to LN i.e, for all values of non-Markovianity present in the channel, QD shows at most one revival for all Haar uniformly generated states irrespective of the content of initial discord and hence the normalised regeneration saturates to unity as shown in Fig. \ref{NormR} for $R_2$ states. Moreover, normalised and mean regeneration of QD for higher value of non-Markovianity (say, \(\alpha \geq 0.5\)), is unity for all random input states having a fixed rank irrespective of initial QD (see Table \ref{tabmeanregDisc}). This means that every state revives after collapse, if, $\alpha \geq 0.5$. Here for a given $\alpha (\geq 0.5$), $\overline{R}_g^D$ remains same with the increase of the rank of the states which is different than the case of entanglement where $\overline{R}_g^E$ decreases with the increase of rank.


		\begin{table}[]
			\resizebox{0.5\textwidth}{!}{\begin{minipage}{0.7\textwidth}
					\caption{$\overline{R_{g}^{D}}$ (Dephasing channel)--Mean regeneration of discord when two-qubit states are passed through single- and double-sided dephasing channels is tabulated for different non-Markovianity parameters. Unlike entanglement, QD of all the randomly generated states becomes nonvanishing after a collapse. } 
					\label{tabmeanregDisc}
					\centering 
					\begin{tabular}{|l|l|l|l|l|l|l|l|}
						\hline
						& \multicolumn{7}{c|}{$\overline{R_{g}^{D}}$}                                                      \\ \hline
						& Rank 1 & \multicolumn{2}{l|}{Rank 2} & \multicolumn{2}{l|}{Rank 3} & \multicolumn{2}{l|}{Rank 4} \\ \hline
						$\alpha$ & double & single       & double       & single       & double       & single       & double       \\ \hline
						&        &              &              &              &              &              &              \\ \hline
						0.2      & 0.988  & 1.000        & 0.979        & 1.000        & 0.949        & 1.000        & 0.898        \\ \hline
						0.3      & 0.996  & 1.000        & 0.998        & 1.000        & 0.999        & 1.000        & 0.997        \\ \hline
						0.5      & 0.999  & 1.000        & 1.000        & 1.000        & 1.000        & 1.000        & 1.000        \\ \hline
						0.6      & 0.999  & 1.000        & 1.000        & 1.000        & 1.000        & 1.000        & 1.000        \\ \hline
						0.7      & 0.999  & 1.000        & 1.000        & 1.000        & 1.000        & 1.000        & 1.000        \\ \hline
						0.8      & 1.000  & 1.000        & 1.000        & 1.000        & 1.000        & 1.000        & 1.000        \\ \hline
						0.9      & 1.000  & 1.000        & 1.000        & 1.000        & 1.000        & 1.000        & 1.000        \\ \hline
					\end{tabular}
			\end{minipage}}
		\end{table}
		
		\item \emph{Discord under double-sided channel.} Moderate presence of non-Markovianity is enough for QD to revive after collapse for almost all the two-qubit random initial states  having  initial QD $\geq 0.5$ which indicates that the minimum amount of QC together with non-Markovianity is able to defeat the destructive effect of noise on states, thereby showing regeneration.
		We find that, unlike single-sided case, QD for low value of $\alpha$, does not always exhibit regeneration (since $\overline{R_{g}^{D}}$ < 1.0). This is possibly   due to fact that the greater amount of noise acts on the state through the double-sided channel. However, with increase of $\alpha$, mean regeneration quickly increases to one -- we refer this fact as \emph{constructive effect of non-Markovian noise} (comparing left and right lower panels of  Fig. \ref{NormR} as well as Fig. \ref{figmeanregent}). Hence, there is a competition between the detrimental effect of noise and constructive response of non-Markovianity -- at low value of $\alpha (\leq 0.3)$, the former overpowers latter and the resulting effect is destructive while for $\alpha (\geq 0.5)$, non-Markovianity defeats the damping  effect of noise, thereby showing constructive impact. \\
		

		Mean regeneration value can reveal  three distinct features  in case of QD which are absent for entanglement  (see  Tables \ref{tabmeanreg} and \ref{tabmeanregDisc} and Fig. \ref{figmeanregent})  -- (1) higher ranked states have same value of mean regeneration than that of  low-ranked random initial states for a fixed \(\alpha\); (2) \(\overline{R}_g^D\) doesn't change with the increase of $\alpha$; (3) double-sided non-Markovian channel leads to a same amount of regeneration as single-sided ones for high non-Markovianity, even though it starts off with less regeneration for low $\alpha$.

	\end{enumerate}

	\subsubsection{Robustness of random states under dephasing noise}

	
	\begin{table}[]
		\resizebox{0.5\textwidth}{!}{\begin{minipage}{0.7\textwidth}
				\caption{ $\overline{p}_{c}^E$ (Dephasing channel)--Mean value of the noise parameter at which entanglement collapses for the first time when two-qubit states are passed through non-Markovian single- and double-sided dephasing channels. It measures the destructive effects of noise on quantum states. } 
				\label{tab_meanpcrit_ent}
				\centering
				\begin{tabular}{|l|l|l|l|l|l|l|l|l|}
					\hline
					\multicolumn{8}{|c|}{$\overline{p}_{c}^E$}                                                                                                                                                                                                                                    \\ \hline
					\multicolumn{1}{|c|}{}         & \multicolumn{1}{c|}{Rank 1}                               & \multicolumn{2}{c|}{Rank 2}                               & \multicolumn{2}{c|}{Rank 3}                               & \multicolumn{2}{c|}{Rank 4}                               \\ \hline
					\multicolumn{1}{|c|}{$\alpha$} & \multicolumn{1}{c|}{double} & \multicolumn{1}{c|}{single} & \multicolumn{1}{c|}{double} & \multicolumn{1}{c|}{single} & \multicolumn{1}{c|}{double} & \multicolumn{1}{c|}{single} & \multicolumn{1}{c|}{double} \\ \hline
					&                             &                             &                             &                             &                             &                             &                             \\ \hline
					0                                                      & 0.261                   & 0.335                   & 0.168                   & 0.224                    & 0.113                    & 0.173                    & 0.084                    \\ \hline
					0.2                                               & 0.226                     & 0.296                    & 0.143                    & 0.196                     & 0.095                    & 0.144                    & 0.07                    \\ \hline
					0.3                                             & 0.211                    & 0.277                    & 0.133                    & 0.182                   & 0.088                    & 0.134                    & 0.065                    \\ \hline
					0.5                                         & 0.185                    & 0.246                    & 0.116                    & 0.16                    & 0.077                    & 0.118                    & 0.056                  \\ \hline
					0.6                                              & 0.174                    & 0.232                    & 0.108                    & 0.152                    & 0.071                    & 0.11                    & 0.053                    \\ \hline
					0.7                                           & 0.164                    & 0.218                    & 0.102                    & 0.142                     & 0.067                    & 0.103                    & 0.049                    \\ \hline
					0.8                                           & 0.155                    & 0.206                    & 0.096                    & 0.134                     & 0.063                    & 0.098                    & 0.046                    \\ \hline
					0.9                                        & 0.146                    & 0.195                    & 0.091                    & 0.127                  & 0.06                    & 0.093                    & 0.044                    \\ \hline
				\end{tabular}
		\end{minipage}}
	\end{table}
	
	To estimate whether QCs of random initial states are robust against noise, we can examine the noise strength that a state can sustain just after passing through the channel. It leads to the value of threshold noise, \(p_c\) of that state. Since we are dealing with non-Markovian channel, $p_c$ should also depend on the non-Markovianity parameter as well as the strength of the noise of the channels acted on a single qubit or both the qubits. Notice that the value of \(p_c\) quantifies the 
	fragile nature of QCs against noise, thereby depicting the harmful consequence of noisy quantum channels. 
	
	\begin{figure}[!ht]
		\resizebox{8.3cm}{5.5cm}{\includegraphics{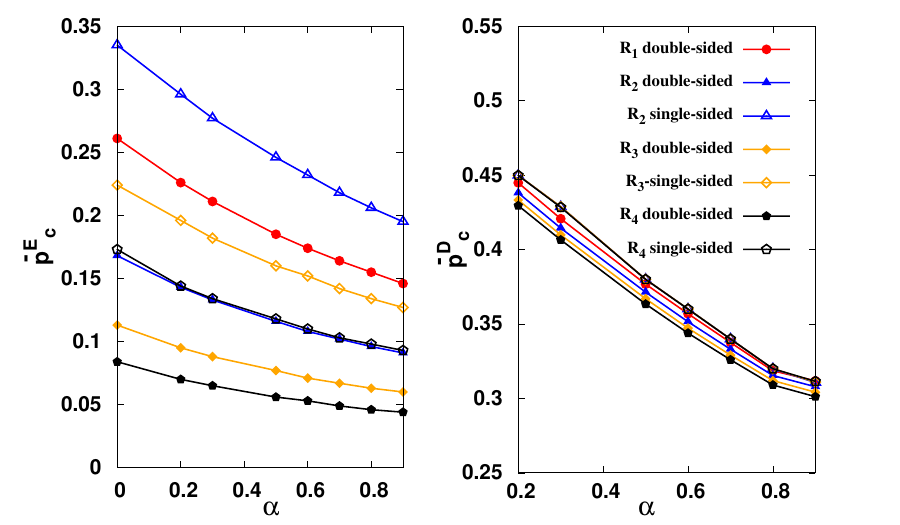}}
		\caption{ (Color online.) Mean critical noise for collapse, $\overline{p}_c^{\mathcal{Q}}$ (\(\mathcal{Q} = E,\, D\)),  (vertical axis)  against $\alpha$ (horizontal axis). Other specifications are same as in Fig. \ref{figmeanregent}. 
		}
		\label{figmeanpcritdephall}
	\end{figure}
	
	As observed before, for a single-sided channel,  LN  as well as  QD of   all pure states do not collapse, thereby having no existence of  \(\overline{p}_c\). For a given QC measure, $\mathcal{Q}$, high values of $\overline{p}_c^Q$ obtained with random input states imply more robustness of $\mathcal{Q}$ on average  against noise.  We find that the pattern of \(\overline{p}_c^E\) for LN  is qualitatively similar when noise acts on  a single side or  both the sides of the randomly generated states -- it posses high value in case of single-sided channel than that of the double-sided ones as clearly shown in Table \ref{tab_meanpcrit_ent} and Fig.  \ref{figmeanpcritdephall}. Interestingly, it decreases with \(\alpha\) for all random states having a fixed rank (see Fig. \ref{figmeanpcritdephall}).  On the other hand, if we fix \(\alpha\), \(\overline{p}_c\) for LN decreases with the rank of the states, thereby indicating robustness of pure states under decoherence.

	The trends of mean critical noise for collapse of QD behave similarly as in entanglement (see right panel of Fig. \ref{figmeanpcritdephall} and Table \ref{tab_meanpcrit_disco}). In general, for a fixed value of \(\alpha\) and fixed rank, it shows high amount of robustness against noise than entanglement which one can expect from the nature of QD itself. Comparing $\overline{p}_c^E$ with $\overline{p}_c^D$ for a fixed $\alpha$ and for a fixed rank, we observe that their difference is on average of the order of $\approx 0.18$ for low rank states while it becomes $\approx 0.25$ for random states with $R_3$ and $R_4$.

	
	\begin{table}[]
		\resizebox{0.5\textwidth}{!}{\begin{minipage}{0.7\textwidth}
				\caption{$\overline{p}_{c}^D$ (Dephasing channel)--Mean value of the noise parameter at which discord collapses for the first time when two-qubit states are passed through non-Markovian single- and double-sided dephasing channels.} 
				\label{tab_meanpcrit_disco}
				\centering
				\begin{tabular}{|l|l|l|l|l|l|l|l|}
					\hline
					& \multicolumn{7}{c|}{$\overline{p_{c}^{D}}$}                                                      \\ \hline
					& Rank 1 & \multicolumn{2}{l|}{Rank 2} & \multicolumn{2}{l|}{Rank 3} & \multicolumn{2}{l|}{Rank 4} \\ \hline
					$\alpha$ & double & single       & double       & single       & double       & single       & double       \\ \hline
					&        &              &              &              &              &              &              \\ \hline
					0        & 0.487  & 0.489        & 0.485        & 0.490        & 0.481        & 0.490        & 0.478        \\ \hline
					0.2      & 0.445  & 0.450        & 0.438        & 0.450        & 0.433        & 0.450        & 0.429        \\ \hline
					0.3      & 0.421  & 0.429        & 0.414        & 0.429        & 0.410        & 0.428        & 0.406        \\ \hline
					0.5      & 0.377  & 0.380        & 0.371        & 0.380        & 0.367        & 0.380        & 0.363        \\ \hline
					0.6      & 0.357  & 0.360        & 0.351        & 0.360        & 0.347        & 0.360        & 0.344        \\ \hline
					0.7      & 0.338  & 0.340        & 0.333        & 0.340        & 0.329        & 0.340        & 0.326        \\ \hline
					0.8      & 0.319  & 0.320        & 0.315        & 0.320        & 0.312        & 0.320        & 0.309        \\ \hline
					0.9      & 0.312  & 0.311        & 0.308        & 0.311        & 0.304        & 0.311        & 0.301        \\ \hline
				\end{tabular}
		\end{minipage}}
	\end{table}
	
	\subsubsection{Mean noise threshold required for quantum correlations to regenerate by random states}
	
	Behavior of mean critical noise required for regeneration of QCs demonstrates that non-Markovian noise is responsible for rebirth of QCs  to happen. Specifically $\overline{p}_{reg}^E$ decreases with the increase of non-Markovianity strength, \(\alpha\) (see Table \ref{tab_meanpreg_ent}).  When noise acts on a single qubit,  LN revives at most once with the increase of \(\alpha\) and the trends of \(\overline{p}_{reg}\) reveals that LN becomes nonvanishing even when the noise strength is very high, like \(p \geq 0.45\) and the lowest noise level in which entanglement revival can be seen is when the presence of non-Markovianity in the channel is  high.  When both the qubits are effected by local non-Markovian dephasing channels, LN also shows revival but the value of \(\overline{p}_{reg}^E\) for random states under double-sided channel is higher than that of the single-sided ones. Both the results possibly pinpoint that the presence of non-Markovianity in channels induces entanglement to resurrect and we note that its impact is more on initial pure states  than the input states with high rank.  
	
	Similar role of non-Markovianity can also be found from the behaviour of  quantum discord which also shows at most one revival after collapse. Like LN, \(\overline{p}_{reg}^D\) decreases with  \(\alpha\) if one fixes the  rank of the states. Moreover, in this case, we find  that unlike \(\overline{p}_{reg}^E\), the difference between \(\overline{p}_{c}^D\) and \(\overline{p}_{reg}^D\) is very low,  and maximum difference can be of the order of \(\approx 0.02-0.03\)  (see Tables \ref{tab_meanpcrit_disco} and \ref{tab_meanpreg_disco}). It implies that  QD revives almost immediately after the first collapse and hence it  can be safely said that  the consequence of  non-Markovianity on QD is more drastic than that of entanglement. 
	As already mentioned,  QD of all states does not collapse. For the states that collapse, we find that the number of states that revive increases significantly with increasing $\alpha$, especially when the noise acts on both the qubits comprising the state (see table \ref{tab_reg_perc_disco}). It further strengthens our claim -- since  the strengths of non-Markovianity is higher when both the sides are passed through the channels compared to the single-sided one,  the  constructive effects of non-Markovianity on QCs in terms of regeneration is more prominent for the former than that of the latter.
	
	\emph{Role of QCs in regeneration.}  Apart from non-Markovianity in channels, QCs possessed by input states also play an important  role in rebirth of QCs. Entire analysis suggests that there exists a minimum value of entanglement, depending on the rank, above which a state has a high possibility  to revive if the channel also posses a moderate amount of non-Markovianity. Specifically, average entanglement required to show regeneration decreases with the increase of non-Markovianity. On the other hand, a certain amount of non-Markovianity in the dephasing channel ensures almost always  QD  to revive and hence the average initial QD required for regeneration in random states is almost constant with  the variation of the rank of the initial state. Such an observation for entanglement and QD is true for both the non-Markovian channels.
	
	\begin{table}[]
		\resizebox{0.5\textwidth}{!}{\begin{minipage}{0.7\textwidth}
				\caption{ $\overline{p}_{reg}^E$ (Dephasing channel)----Mean value of the noise parameter at which entanglement revives for the first time when two-qubit states are passed through non-Markovian single- and double-sided dephasing channels. This is a complementary quantity of \(\overline{p}_c^E \) and can capture the effect of non-Markovianity on states.} 
				\label{tab_meanpreg_ent}
				\centering
				\begin{tabular}{|l|l|l|l|l|l|l|l|l|}
					\hline
					\multicolumn{8}{|c|}{$\overline{p}_{reg}^E$}                                                                                                                                                                                                                                    \\ \hline
					\multicolumn{1}{|c|}{}         & \multicolumn{1}{c|}{Rank 1}                               & \multicolumn{2}{c|}{Rank 2}                               & \multicolumn{2}{c|}{Rank 3}                               & \multicolumn{2}{c|}{Rank 4}                               \\ \hline
					\multicolumn{1}{|c|}{$\alpha$} & \multicolumn{1}{c|}{double} & \multicolumn{1}{c|}{single} & \multicolumn{1}{c|}{double} & \multicolumn{1}{c|}{single} & \multicolumn{1}{c|}{double} & \multicolumn{1}{c|}{single} & \multicolumn{1}{c|}{double} \\ \hline
					&                             &                             &                             &                             &                             &                             &                             \\ \hline
					0.2                                             & 0.495                    & 0.484                    & -                    & 0.493                   & -                    & 0.494                    & -                    \\ \hline
					0.3                                             & 0.489                    & 0.474                    & -                    & 0.485                   & -                    & 0.491                    & -                    \\ \hline
					0.5                                         & 0.478                    & 0.449                    & 0.49                    & 0.47                    & 0.5                    & 0.48                    & -                  \\ \hline
					0.6                                              & 0.474                    & 0.437                    & 0.487                    & 0.462                    & 0.493                    & 0.475                    & -                    \\ \hline
					0.7                                           & 0.47                    & 0.424                    & 0.485                    & 0.454                     & 0.495                    & 0.467                    & -                    \\ \hline
					0.8                                           & 0.469                    & 0.412                    & 0.482                    & 0.446                     & 0.489                    & 0.462                    & 0.498                    \\ \hline
					0.9                                        & 0.466                    & 0.399                    & 0.48                    & 0.437                  & 0.486                    & 0.454                    & 0.489                    \\ \hline
				\end{tabular}
		\end{minipage}}
	\end{table}
	
	\begin{table}[]
		\resizebox{0.5\textwidth}{!}{\begin{minipage}{0.7\textwidth}
				\caption{$\overline{p}_{reg}^{D}$  (Dephasing channel)----Mean value of the noise parameter at which QD revives for the non-Markovian single- and double-sided dephasing channels. It decreases with \(\alpha\), thereby showing that the non-Markovianity  can induce regeneration of QCs. } 
				\label{tab_meanpreg_disco}
				\centering
				\begin{tabular}{|l|l|l|l|l|l|l|l|}
					\hline
					& \multicolumn{7}{c|}{$\overline{p_{reg}^{D}}$}                                                    \\ \hline
					& Rank 1 & \multicolumn{2}{l|}{Rank 2} & \multicolumn{2}{l|}{Rank 3} & \multicolumn{2}{l|}{Rank 4} \\ \hline
					$\alpha$ & double & single       & double       & single       & double       & single       & double       \\ \hline
					&        &              &              &              &              &              &              \\ \hline
					0.2      & 0.466  & 0.460        & 0.473        & 0.460        & 0.477        & 0.460        & 0.480        \\ \hline
					0.3      & 0.443  & 0.440        & 0.449        & 0.440        & 0.454        & 0.440        & 0.457        \\ \hline
					0.5      & 0.397  & 0.390        & 0.403        & 0.390        & 0.408        & 0.390        & 0.411        \\ \hline
					0.6      & 0.376  & 0.370        & 0.382        & 0.370        & 0.386        & 0.370        & 0.390        \\ \hline
					0.7      & 0.356  & 0.350        & 0.362        & 0.350        & 0.366        & 0.350        & 0.370        \\ \hline
					0.8      & 0.340  & 0.336        & 0.344        & 0.335        & 0.348        & 0.336        & 0.351        \\ \hline
					0.9      & 0.331  & 0.322        & 0.335        & 0.322        & 0.339        & 0.323        & 0.342        \\ \hline
				\end{tabular}
		\end{minipage}}
	\end{table}

	\section{Random states passed through depolarizing Channel}
	\label{sec_depol}
	
	Let us move to a scenario where a single qubit or both the qubits of the input states with different ranks are sent through depolarizing non-Markovian channels having different strengths of non-Markovianity. We will show that although non-Markovian effect on QCs obtained for dephasing and depolarizing channels are of similar kind, certain differences in the behaviour of QCs are also present. Let us begin our discussion with some analytical derivations. 
	
	\textbf{Proposition 4.} \textit{A pure two-qubit state, when passed through the single-sided depolarising channel, shows no regeneration of entanglement while the constructive response of non-Markovianity ensures revival in case of the double-sided channel.}\\
	
\begin{figure}[!ht]
		\resizebox{8cm}{5cm}{\includegraphics{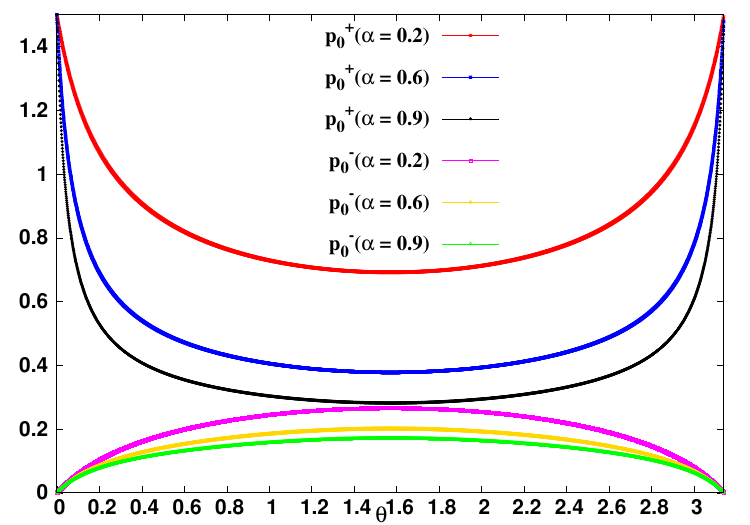}}
		\caption{(Color online. ) Plot of $p_0^{\pm}$ (ordinate) against $\theta$ (abscissa) for different values of $\alpha$. We show $p_0^+$ in the three upper lines for $\alpha = 0.2$ (topmost), $\alpha = 0.6$ (second from the top) and $\alpha = 0.9$ (third from the top). The bottom three lines represent $p_0^-$ for $\alpha = 0.2$ (fourth from the top), $\alpha = 0.6$ (fifth from the top) and $\alpha = 0.9$ (bottom-most).
		}
		\label{prop4}
	\end{figure}

	\textit{Proof.} After the action of the non-Markovian depolarising channel on \(|\psi_{AB} \rangle =\cos \theta/2 |00\rangle + \sin \theta/2 |11\rangle\), the negativity of the final state $\rho_f = \sum_{i,j} (K_i^{dp} \otimes K_j^{dp}) |\psi_{AB}\rangle \langle\psi_{AB}| (K_i^{dp} \otimes K_j^{dp})^\dagger$ 
	reads as
	\begin{eqnarray}
		&& |\frac{2p(3\alpha p (p - 1) - 1)}{6 (-1 + 3 \alpha (-1 + p)^2 p)} \nonumber \\
		&&+ \frac{\sqrt{p^2(1 - 3\alpha p (p - 1))^2(3 + \cos2\theta)-f(\alpha, p) \sin^2\theta}}{6 (-1 + 3 \alpha (-1 + p)^2 p)}|\nonumber\\
		\label{r1entdepo}
	\end{eqnarray}
	where $f(\alpha,p) = -9 + p (24 - 14 p + 3 \alpha (-1 + p) (-18 + p (18 + 4 p + 3 \alpha (-1 + p) (-9 + (-6 + p) p))))$.  Analyzing Eq. (\ref{r1entdepo}) for the entire range of  $p$ and for fixed values of $\alpha$ and $\theta$, we find no revival of entanglement, i.e., we find that the above equation  has only one zero in terms of \(p\). E.g., setting $\alpha = 0.9$ and $\theta = 0.6$, the only time the function vanishes is at $p = 0.219$ and thereafter, entanglement remains zero.\\
	On the other hand, in the case of double-sided depolarising channel, we can again obtain a cumbersome expression for negativity. In this situation, ignoring terms of $\mathcal{O}(p^3)$,  the entanglement goes to zero at the damping parameters (in terms of the other parameters),  
	\begin{eqnarray}
		p_0^\pm = \frac{6 + 3 (4 + 9 \alpha) \sin\theta}{(8 + (16 + 27 \alpha (4 + 3 \alpha)) \sin\theta)}  \nonumber\\
		\pm \frac{3 \sqrt{2} \sqrt{2 - 9 \alpha + 9 \alpha \cos2 \theta + 2 (2 + 9 \alpha) \sin\theta}}{(8 + (16 + 27 \alpha (4 + 3 \alpha)) \sin\theta)}, 
	\end{eqnarray}
	which we plot in Fig. \ref{prop4} by varying \(\theta\) for different values of \(\alpha\).
	In this case,  we find that $p_0^-$ is always a valid damping parameter while $p_0^+ <=1$ only for $\alpha \geq \alpha_c$  where $\alpha_c $ can be found  for a given \(\theta\). Notice first that for $\alpha = 0$, which denotes the Markovian regime, entanglement does not show any regeneration. For $\alpha < \alpha_c$, we get a valid noise parameter as  $p_0^-$, which implies that the entanglement undergoes collapse only. For $\alpha \geq \alpha_c$,  both $p_0^-$ and \(p_0^+\) are valid which definitely conclude that entanglement collapses twice, thereby exhibiting regeneration in between these two collapses. \\
	Another confirmation for the same is that, both the left hand and right hand limits around $p_0^-$ are positive, having the value $6h \sqrt{2} \sqrt{2 - 9 \alpha + 9 \alpha \cos 2 \theta + 2 (2 + 9 \alpha) \sin\theta}$ at both $p_0^- + h$ and $p_0^- - h$ as $h \to 0$ ignoring $\mathcal{O}(p^3)$. It also demonstrates the constructive effect of non-Markovianity, since the higher amount of noise in the double-sided channel helps in the regeneration. E.g.,  $\alpha = 0.9$ and $\theta = 0.6$, we find $p_c = 0.131$,  and $0.355$.  \hfill $\blacksquare$  \\

	The above results clearly establish the constructive reaction of non-Markovianity on QCs, thereby overcoming the destructive effects of noise on systems. As mentioned before, when double-sided channels is active on states, noise as well as non-Markovianity both increase and hence for a high value of \(\alpha\), non-Markovianity wins which is responsible for revival of QCs.  Let us now concentrate on the numerically obtained observations which confirm the above results, irrespective of the rank of the two-qubit states. 
	
	\begin{table}[]
		\resizebox{0.5\textwidth}{!}{\begin{minipage}{0.7\textwidth}
				\caption{$Regeneration \: \% $  (Dephasing channel)--Percentage of states which undergo regeneration when two-qubit states of different ranks are passed through singe- and double-sided non-Markovian dephasing channel} 
				\label{tab_reg_perc_disco}
				\centering
				\begin{tabular}{|l|l|l|l|l|l|l|l|}
					\hline
					& \multicolumn{7}{c|}{$Regeneration \: \%$}                                                        \\ \hline
					& Rank 1 & \multicolumn{2}{l|}{Rank 2} & \multicolumn{2}{l|}{Rank 3} & \multicolumn{2}{l|}{Rank 4} \\ \hline
					$\alpha$ & double & single       & double       & single       & double       & single       & double       \\ \hline
					&        &              &              &              &              &              &              \\ \hline
					0.2      & 98.820 & 99.996       & 97.942       & 100.000      & 94.906       & 100.000      & 89.808       \\ \hline
					0.3      & 99.614 & 100.000      & 99.878       & 100.000      & 99.882       & 100.000      & 99.978       \\ \hline
					0.5      & 99.916 & 100.000      & 99.996       & 100.000      & 99.996       & 100.000      & 99.999       \\ \hline
					0.6      & 91.944 & 100.000      & 100.000      & 100.000      & 100.000      & 100.000      & 100.000      \\ \hline
					0.7      & 99.955 & 100.000      & 100.000      & 100.000      & 100.000      & 100.000      & 100.000      \\ \hline
					0.8      & 99.977 & 100.000      & 100.000      & 100.000      & 100.000      & 100.000      & 100.000      \\ \hline
					0.9      & 99.984 & 100.000      & 100.000      & 100.000      & 100.000      & 100.000      & 100.000      \\ \hline
				\end{tabular}
		\end{minipage}}
	\end{table}
	
	\begin{figure}[!ht]
		\resizebox{9.0cm}{6cm}{\includegraphics{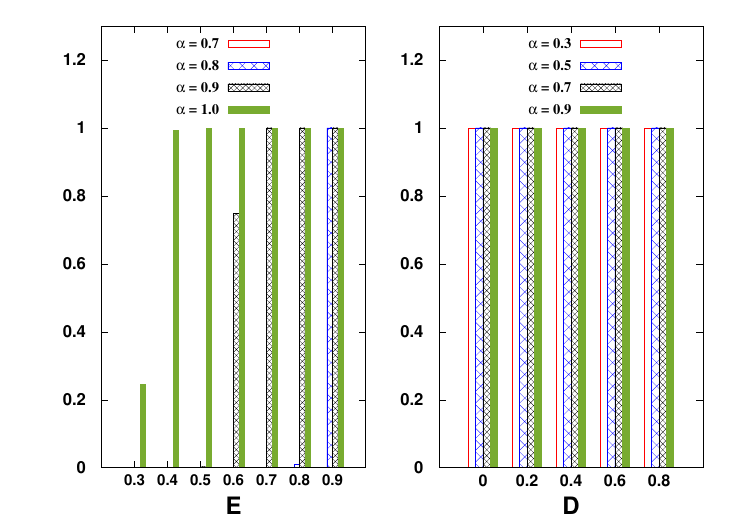}}
		\caption{(Color online. ) Plot of the distribution of normalised regeneration, $R_g^{N}$, against initial LN (left) and QD (right).  Both the qubits of random rank-2 states are sent through the depolarizing non-Markovian channel with different non-Markovian parameters, \(\alpha\). Other specifications are same as in Fig. \ref{NormR}.
		}
		\label{NormR_depo}
	\end{figure}

	\begin{table}[]
		\resizebox{0.5\textwidth}{!}{\begin{minipage}{0.7\textwidth}
				\caption{ $\overline{R}_{g}^{E}$  (Depolarizing channel)--Mean regeneration of entanglement when two-qubit states are passed through single- and double-sided depolarising channel is tabulated for different non-Markovianity parameters. Notice that after the action of single-sided channel,  there is no regeneration of entanglement. Constructive feedback from non-Markovianity is visible here. Since for the double-sided case, noise and non-Markovianity both increase, non-Markovianity can sometimes suppress the damping  nature of noise. } 
				\centering
				\begin{tabular}{|l|l|l|l|l|l|l|l|l|}
					\hline
					\multicolumn{5}{|c|}{$\overline{R}_{g}^E$}                                                                                                                                                                                                                                    \\ \hline
					\multicolumn{1}{|c|}{}         & \multicolumn{1}{c|}{Rank 1}                               & \multicolumn{1}{c|}{Rank 2}                               & \multicolumn{1}{c|}{Rank 3}                               & \multicolumn{1}{c|}{Rank 4}                               \\ \hline
					\multicolumn{1}{|c|}{$\alpha$}  & \multicolumn{1}{c|}{double}  & \multicolumn{1}{c|}{double} & \multicolumn{1}{c|}{double}  & \multicolumn{1}{c|}{double} \\ \hline
					&                             &                             &                             &                             \\ \hline
					0.3                                        & 0                    & 0                    & 0                    & 0                    \\ \hline
					0.7                                        & 0                    & 0                    & 0                    & 0                    \\ \hline
					0.8                                        & 0.128                    & 0.001                    & 0                    & 0                    \\ \hline
					0.9                                        & 0.579                     & 0.118                    & 0.016                    & 0.002                    \\ \hline
					1                                          & 0.843                    & 0.484                   & 0.198                    & 0.081                    \\ \hline
				\end{tabular}
		\end{minipage}}
	\end{table}
	
	\begin{table}[]
		\resizebox{0.5\textwidth}{!}{\begin{minipage}{0.7\textwidth}
				\caption{ $\overline{R}_{g}^{D}$  (Depolarizing channel)--Mean regeneration of discord when two-qubit states are passed through single- and double-sided depolarising channel is tabulated for different non-Markovianity parameters. Again it increases with non-Markovianity, especially for the double-sided channels, thereby highlighting the  fact that non-Markovianity is responsible for resurrections. } 
				\label{tab_rg_depo}
				\centering
				\begin{tabular}{|l|l|l|l|l|l|l|l|l|}
					\hline
					& \multicolumn{8}{c|}{$\overline{R_{g}^{D}}$}                                                                           \\ \hline
					& \multicolumn{2}{l|}{Rank 1} & \multicolumn{2}{l|}{Rank 2} & \multicolumn{2}{l|}{Rank 3} & \multicolumn{2}{l|}{Rank 4} \\ \hline
					$\alpha$ & single       & double       & single       & double       & single       & double       & single       & double       \\ \hline
					&              &              &              &              &              &              &              &              \\ \hline
					0        & 1.000        & 0.998        & 1.000        & 0.999        & 1.000        & 0.999        & 1.000        & 0.999        \\ \hline
					0.2      & 1.000        & 0.998        & 1.000        & 0.999        & 1.000        & 0.999        & 1.000        & 0.999        \\ \hline
					0.3      & 1.000        & 0.998        & 1.000        & 0.999        & 1.000        & 0.999        & 1.000        & 0.999        \\ \hline
					0.5      & 1.000        & 0.999        & 1.000        & 1.001        & 1.000        & 1.000        & 1.000        & 1.000        \\ \hline
					0.6      & 1.000        & 1.003        & 1.000        & 1.010        & 1.000        & 1.004        & 1.000        & 1.004        \\ \hline
					0.7      & 1.000        & 1.015        & 1.000        & 1.041        & 0.999        & 1.028        & 0.999        & 1.028        \\ \hline
					0.8      & 1.000        & 1.048        & 1.000        & 1.124        & 1.000        & 1.135        & 1.000        & 1.161        \\ \hline
					0.9      & 1.000        & 1.173        & 1.000        & 1.385        & 1.000        & 1.518        & 1.000        & 1.630        \\ \hline
				\end{tabular}
		\end{minipage}}
	\end{table}
	
	\subsubsection{Single- vs. double-sided non-Markovian channels}
	
	We first describe how non-Markovianity activates rebirth of entanglement  in presence of high amount of noise and then we move to quantum discord in a similar situation.  \\
	\begin{enumerate}
		
		\item \emph{Entanglement under depolarizing non-Markovian channel.} Let us first consider the scenario when noise only acts on a single side.  Interestingly, we find that LN  of the resulting states from random input states  does not show any revival after collapse for any rank and for any non-Markovian strength of the channel. As we discussed before,  although the presence of non-Markovianity causes entanglement  to revive,  in this scenario,  strength of non-Markovianity is possibly not enough to overcome the power of noise, thereby showing  destructive effects of depolarizing channel on entanglement. \\
		
		On the contrary, when both the qubits are passed through a local depolarizing channel, entanglement resurrects  for high value of $\alpha$, say $0.8$ and above, as depicted in left panel of Fig. \ref{NormR_depo} for random rank-2 input states.  As we have seen for QD with dephasing channel,  we here also report a \emph{constructive} effect of noise on entanglement -- more noise on the state shows possibility of revival  while less noisy states do not. It possibly shows that  the effect of damping parameter $p$ which, in general,  destroys QCs  can be overcome by the non-Markovianity,  $\alpha$, that tends to shield  QCs from noise. Hence when non-Markovian effect exceeds the damping effect, QCs, either in the form of entanglement or QD, revives. We also notice that to show revival, random initial states also should posses high amount of entanglement on average. 
		We find that \(\overline{R}_g^E\) shows similar trend like dephasing channel, i.e. it increases with the non-Markovianity for a given rank of random states, thereby confirming importance of non-Markovianity for revival of entanglement. 
		
		\begin{table}[]
			\resizebox{0.5\textwidth}{!}{\begin{minipage}{0.7\textwidth}
					\caption{$Regeneration \: \%$  (Depolarizing channel)--Percentage of states which undergo regeneration when two-qubit states of different ranks are passed through singe- and double-sided non-Markovian depolarising channel} 
					\label{tab_reg_perc_disc_depo}
					\centering
					\begin{tabular}{|l|l|l|l|l|l|l|l|l|}
						\hline
						& \multicolumn{8}{c|}{$Regeneration \%$}                                                                                \\ \hline
						& \multicolumn{2}{l|}{Rank 1} & \multicolumn{2}{l|}{Rank 2} & \multicolumn{2}{l|}{Rank 3} & \multicolumn{2}{l|}{Rank 4} \\ \hline
						$\alpha$ & single        & double      & single        & double      & single        & double      & single        & double      \\ \hline
						&               &             &               &             &               &             &               &             \\ \hline
						0        & 100.000       & 99.820      & 100.000       & 99.914      & 100.000       & 99.992      & 100.000       & 99.986      \\ \hline
						0.2      & 100.000       & 9.822       & 100.000       & 99.948      & 100.000       & 99.998      & 100.000       & 99.996      \\ \hline
						0.3      & 100.000       & 99.810      & 100.000       & 99.966      & 100.000       & 99.986      & 100.000       & 99.994      \\ \hline
						0.5      & 100.000       & 99.822      & 100.000       & 99.956      & 100.000       & 99.994      & 100.000       & 99.992      \\ \hline
						0.6      & 99.980        & 99.834      & 100.000       & 99.950      & 100.000       & 99.986      & 100.000       & 99.994      \\ \hline
						0.7      & 100.000       & 99.848      & 100.000       & 99.940      & 100.000       & 99.992      & 100.000       & 99.994      \\ \hline
						0.8      & 100.000       & 99.866      & 100.000       & 99.940      & 100.000       & 99.992      & 100.000       & 99.996      \\ \hline
						0.9      & 100.000       & 99.842      & 100.000       & 99.940      & 100.000       & 99.998      & 100.000       & 99.994      \\ \hline
					\end{tabular}
			\end{minipage}}
		\end{table}
		
		\item \emph{QD under depolarizing channel.} Like dephasing channel,  after passing  through a noisy channel (either a single qubit or both the qubits)  having moderate value of non-Markovianity, QD of the output  states  always revives with the increase of $p$ irrespective of the rank of the initial states, and the amount of QD in the initial state. Unlike dephasing channel, for a fixed non-Markovianity, the number of  regeneration seen for depolarizing channel is higher than that of dephasing one. (see  right part of Fig. \ref{NormR_depo} for random rank-2 states).  
		For single- as well as double-sided cases, \(\overline{R}_g^D\) increases with non-Markovianity for all the ranks of the random states (see Table \ref{tab_rg_depo}). Furthermore, we find that even though \(\overline{R}_g^D\) is less for double-sided channel action than the single-sided case for low values of $\alpha$ ($\leq$ 0.4), it surpasses the latter significantly, as we move on to stronger non-Markovian regimes. The constructive feedback of non-Markovianity is apparent in this case (see Table \ref{tab_reg_perc_disc_depo}).
		\begin{table}[]
			\resizebox{0.5\textwidth}{!}{\begin{minipage}{0.7\textwidth}
					\caption{$\overline{p}_{c}^{E}$  (Depolarizing channel)--Mean value of the noise parameter at which entanglement collapses for the first time with single- and double-sided depolarising channels.} 
					\label{tab_meanpcrit_ent_depo}
					\centering
					\begin{tabular}{|l|l|l|l|l|l|l|l|l|}
						\hline
						\multicolumn{9}{|c|}{$\overline{p}_{c}^E$}                                                                                                                                                                                                                                    \\ \hline
						\multicolumn{1}{|c|}{}         & \multicolumn{2}{c|}{Rank 1}                               & \multicolumn{2}{c|}{Rank 2}                               & \multicolumn{2}{c|}{Rank 3}                               & \multicolumn{2}{c|}{Rank 4}                               \\ \hline
						\multicolumn{1}{|c|}{$\alpha$} & \multicolumn{1}{c|}{single} & \multicolumn{1}{c|}{double} & \multicolumn{1}{c|}{single} & \multicolumn{1}{c|}{double} & \multicolumn{1}{c|}{single} & \multicolumn{1}{c|}{double} & \multicolumn{1}{c|}{single} & \multicolumn{1}{c|}{double} \\ \hline
						&                             &                             &                             &                             &                             &                             &                             &                             \\ \hline
						0                              & 0.491                         & 0.227                    & 0.306                    & 0.156                    & 0.21                    & 0.108                    & 0.159                    & 0.082                    \\ \hline
						0.2                            & 0.36                    & 0.15                     & 0.209                    & 0.101                    & 0.139                     & 0.068                    & 0.104                    & 0.051                    \\ \hline
						0.3                            & 0.3                    & 0.126                    & 0.177                    & 0.084                    & 0.117                    & 0.057                    & 0.088                    & 0.042                    \\ \hline
						0.5                            & 0.23                    & 0.094                    & 0.132                    & 0.063                    & 0.087                    & 0.042                    & 0.065                    & 0.031                    \\ \hline
						0.6                            & 0.2                     & 0.083                    & 0.117                    & 0.055                    & 0.077                    & 0.037                    & 0.057                    & 0.027                    \\ \hline
						0.7                            & 0.18                    & 0.074                    & 0.104                    & 0.049                    & 0.069                     & 0.032                    & 0.051                    & 0.024                    \\ \hline
						0.8                            & 0.16                    & 0.071                    & 0.095                    & 0.044                    & 0.062                     & 0.029                    & 0.046                    & 0.021                    \\ \hline
						0.9                            & 0.14                    & 0.068                    & 0.086                    & 0.040                    & 0.057                    & 0.026                    & 0.042                    & 0.019                    \\ \hline
					\end{tabular}
			\end{minipage}}
		\end{table}
		
		\begin{table}[]
			\resizebox{0.5\textwidth}{!}{\begin{minipage}{0.7\textwidth}
					\caption{$\overline{p}_{c}^{D}$  (Depolarizing channel)--Mean value of the noise parameter at which discord collapses for the first time for the depolarising channel.} 
					\label{tab_meanpcrit_disc_depo}
					\centering
					\begin{tabular}{|l|l|l|l|l|l|l|l|l|}
						\hline
						& \multicolumn{8}{c|}{$\overline{p_{c}^{D}}$}                                                                           \\ \hline
						& \multicolumn{2}{l|}{Rank 1} & \multicolumn{2}{l|}{Rank 2} & \multicolumn{2}{l|}{Rank 3} & \multicolumn{2}{l|}{Rank 4} \\ \hline
						$\alpha$ & single       & double       & single       & double       & single       & double       & single       & double       \\ \hline
						&              &              &              &              &              &              &              &              \\ \hline
						0        & 0.750        & 0.693        & 0.747        & 0.678        & 0.747        & 0.674        & 0.746        & 0.669        \\ \hline
						0.2      & 0.608        & 0.545        & 0.603        & 0.529        & 0.602        & 0.524        & 0.600        & 0.520        \\ \hline
						0.3      & 0.523        & 0.470        & 0.520        & 0.456        & 0.520        & 0.452        & 0.519        & 0.448        \\ \hline
						0.5      & 0.390        & 0.353        & 0.389        & 0.343        & 0.389        & 0.340        & 0.389        & 0.337        \\ \hline
						0.6      & 0.339        & 0.311        & 0.338        & 0.302        & 0.339        & 0.300        & 0.339        & 0.297        \\ \hline
						0.7      & 0.299        & 0.278        & 0.298        & 0.270        & 0.299        & 0.267        & 0.299        & 0.265        \\ \hline
						0.8      & 0.269        & 0.250        & 0.269        & 0.243        & 0.270        & 0.241        & 0.270        & 0.239        \\ \hline
						0.9      & 0.258        & 0.239        & 0.258        & 0.232        & 0.259        & 0.230        & 0.259        & 0.229        \\ \hline
					\end{tabular}
			\end{minipage}}
		\end{table}

		\item \emph{Mean critical noise for collapse. } As defined in Sec. \ref{sec_signi}, higher value of mean critical noise for collapse indicates  states to be  more robust against noise.
		Like dephasing channel, the trends of $\overline{p}_c^E$ is the same for both single - and double-sided channels i.e. $\overline{p}_{c}^E$ for LN  decreases with increasing $\alpha$ for all random states having different ranks (see Table \ref{tab_meanpcrit_ent_depo}). However, for a fixed rank and fixed non-Markovian strength, $\overline{p}_{c}^E$ obtained for LN  is always greater when a single qubit of the state is noisy than the case for double-sided channel, while it decreases with rank  for a given value of \(\alpha\). As depicted in Fig.  \ref{figmeanpcr_depo_ent}, random pure states whose single qubit is affected by noise have a special status -- they are maximally robust against noise -- in worst case, the difference of \(\overline{p}_c^E\) with other rank states is of the order of $\approx 0.05$ among all other randomly simulated states with different ranks, irrespective of the strength of  non-Markovianity. \\
		The behavior of \(\overline{p}_c^D\) is exactly similar to that of entanglement (see Table \ref{tab_meanpcrit_disc_depo}). In particular, for random states with a fixed rank, it decreases with the increase of \(\alpha\) (see Fig. \ref{figmeanpcr_depo_ent}). Clearly, it supports the fact that non-Markovianity can induce regeneration or nonmonotonicity in QC but can decrease sustainability of initial QCs.
		
		\begin{figure}[!ht]
			\resizebox{9cm}{7cm}{\includegraphics{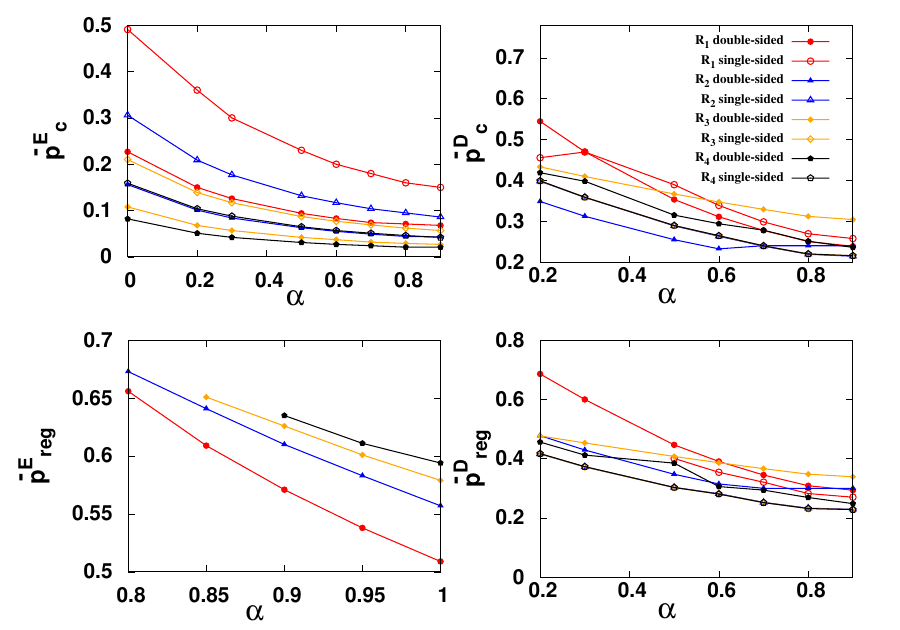}}
			\caption{(Color online.) Upper Panel: Mean critical noise for collapse, $\overline{p}_c^{\mathcal{Q}}$,  vs. \(\alpha \) for depolarizing channel. Lower panel: Behavior of $\overline{p}_{reg}^{\mathcal{Q}}$  with  $\alpha$. Other specifications are same as in Fig. \ref{figmeanregent}.
			}
			\label{figmeanpcr_depo_ent}
		\end{figure}
		
		\begin{table}[]
			\resizebox{0.5\textwidth}{!}{\begin{minipage}{0.7\textwidth}
					\caption{$\overline{p}_{reg}^{E}$  (Depolarizing channel)--Mean value of the noise parameter at which entanglement regenerates for random states sent through the  non-Markovian single- and double-sided depolarising channels. It decreases with non-Markovianity which possibly shows that the non-Markovianity in the channel does not increase robustness. } 
					\label{tab_meanpreg_ent_depo}
					\centering
					\begin{tabular}{|l|l|l|l|l|l|l|l|l|}
						\hline
						\multicolumn{5}{|c|}{$\overline{p}_{reg}^E$}                                                                                                                                                                                                                                    \\ \hline
						\multicolumn{1}{|c|}{}         & \multicolumn{1}{c|}{Rank 1}                               & \multicolumn{1}{c|}{Rank 2}                               & \multicolumn{1}{c|}{Rank 3}                               & \multicolumn{1}{c|}{Rank 4}                               \\ \hline
						\multicolumn{1}{|c|}{$\alpha$}  & \multicolumn{1}{c|}{double}  & \multicolumn{1}{c|}{double} & \multicolumn{1}{c|}{double}  & \multicolumn{1}{c|}{double} \\ \hline
						&                             &                             &                             &                             \\ \hline
						0.8                                        & 0.656                    & 0.673                    & 0                    & 0                    \\ \hline
						0.85                                        & 0609                    & 0.641                    & 0.651                    & 0                    \\ \hline
						0.9                                      & 0.571                    & 0.61                    & 0.626                    & 0.635                    \\ \hline
						0.95                                        & 0.538                     & 0.583                    & 0.601                    & 0.611                    \\ \hline
						1                                          & 0.509                    & 0.557                   & 0.579                    & 0.594                    \\ \hline
					\end{tabular}
			\end{minipage}}
		\end{table}

		
		
		
		\begin{table}[]
			\resizebox{0.5\textwidth}{!}{\begin{minipage}{0.7\textwidth}
					\caption{$\overline{p}_{reg}^{D}$  (Depolarizing channel)--Mean value of the noise parameter at which discord revives for the first time for the depolarizing channel. } 
					\label{tabmeanpreg_disco_depo}
					\centering
					\begin{tabular}{|l|l|l|l|l|l|l|l|l|}
						\hline
						& \multicolumn{8}{c|}{$\overline{p_{reg}^{D}}$}                                                                         \\ \hline
						& \multicolumn{2}{l|}{Rank 1} & \multicolumn{2}{l|}{Rank 2} & \multicolumn{2}{l|}{Rank 3} & \multicolumn{2}{l|}{Rank 4} \\ \hline
						$\alpha$ & single       & double       & single       & double       & single       & double       & single       & double       \\ \hline
						&              &              &              &              &              &              &              &              \\ \hline
						0        & 0.760        & 0.816        & 0.763        & 0.832        & 0.763        & 0.836        & 0.764        & 0.841        \\ \hline
						0.2      & 0.621        & 0.686        & 0.623        & 0.707        & 0.623        & 0.713        & 0.623        & 0.719        \\ \hline
						0.3      & 0.541        & 0.600        & 0.543        & 0.620        & 0.542        & 0.625        & 0.542        & 0.631        \\ \hline
						0.5      & 0.401        & 0.447        & 0.403        & 0.461        & 0.403        & 0.465        & 0.404        & 0.469        \\ \hline
						0.6      & 0.354        & 0.391        & 0.356        & 0.402        & 0.355        & 0.405        & 0.356        & 0.408        \\ \hline
						0.7      & 0.320        & 0.345        & 0.321        & 0.355        & 0.320        & 0.357        & 0.320        & 0.360        \\ \hline
						0.8      & 0.282        & 0.309        & 0.284        & 0.317        & 0.282        & 0.319        & 0.283        & 0.322        \\ \hline
						0.9      & 0.270        & 0.293        & 0.271        & 0.301        & 0.270        & 0.304        & 0.271        & 0.306        \\ \hline
					\end{tabular}
			\end{minipage}}
		\end{table}

		\item \emph{Mean critical noise for revival.} In case of entanglement, since there is no regeneration when a single qubit is sent via noisy channel, there does not exist any non-trivial value of  mean critical noise for revival. If both the qubits are passed through the channel, there is regeneration for higher values of non-Markovianity, and \(\overline{p}_{reg}^E\) decreases with non-Markovianity (see Table \ref{tab_meanpreg_ent_depo}). On the other hand, the behaviour of \(\overline{p}_{reg}^D\) is identical with the dephasing channel (see  Fig. \ref{figmeanpcr_depo_ent} and Table \ref{tabmeanpreg_disco_depo}).
		
		\emph{Activation of rebirth of QCs. } Let us check how non-Markovianity can activate rebirth of  QCs  after the first collapse. To estimate this,  we consider the difference between $\overline{p}_c^{\mathcal{Q}}$ and $\overline{p}_{reg}^{\mathcal{Q}}$ for a fixed rank of the input state and for a fixed \(\alpha\), i.e we evaluate \(\delta^{\mathcal{Q}}_{rb} = \overline{p}_{reg}^{\mathcal{Q}} - \overline{p}_{c}^{\mathcal{Q}}\) by comparing  Tables  \ref{tab_meanpcrit_ent}, \ref{tab_meanpreg_ent}, \ref{tab_meanpcrit_ent_depo} and \ref{tab_meanpreg_ent_depo} for LN and Tables \ref{tab_meanpcrit_disco}, \ref{tab_meanpreg_disco}, \ref{tab_meanpcrit_disc_depo} and \ref{tabmeanpreg_disco_depo}) for QD.  
		
		\begin{itemize}
			
			\item In case of a single-sided dephasing channel, $0.188 \lesssim \delta^{E}_{rb}\lesssim 0.365$, while for the double-sided case, it lies between $0.269$ and $0.452$. On the other hand, for depolarizing double-sided channel,  it lies between $0.585$ and $0.616$. The high gap between collapse and revival again establishes that resurrection of entanglement after collapse is possibly difficult even in presence of non-Markovianity. 
			
			\item  In case of QD, for a single- and double- sided dephasing channels, we respectively get $0.01\lesssim \delta^{D}_{rb} \lesssim 0.016$ and  $0.018\lesssim \delta^{D}_{rb}\lesssim 0.05$.  For a double-sided depolarizing channel, the maximum is $0.19$ and the minimum is $0.055$ while  $0.011\lesssim\delta^{D}_{rb}\lesssim 0.023$ for a single-sided depolarizing one. The small gap shown by QD illustrates that non-Markovianity possibly helps QD to overcome the barrier of noise in the channel more strongly compared to entanglement. 
			
		\end{itemize}

		
	\end{enumerate}

	\emph{Remark.} The form of  the depolarizing map ensures that the damping effects are present from all the directions and hence it  destroys quantum correlations very fast, thereby showing a very low critical noise for collapse. 
	When a single qubit of a two-qubit state is effected by the non-Markovian depolarizing channel, the maximum allowed value of non-Markovianity which is typically responsible for resurrection is possibly not enough to overcome the damping effects, thereby showing absence of  revival while  when  both  the  qubits  are  sent  through  the  depolarizing channel,  both  the  damping  and  non-Markovian  effects  are  strong  and  for  high  values  of non-Markovianity, it dominates over the detrimental effects of noise. On the other hand, in case of dephasing channels, since the noise acts only in the \(z\)-direction and hence  its destructive effect on states is not so pronounced, the competition between damping and non-Markovianity is comparatively weaker than that of the depolarizing channel.

	\section{Conclusion}
	\label{sec_conclu}
	
	
	The distribution of a certain property seen in random quantum states can, in general, be different than the one observed for a specific
	class of states although these random states can also possess an underlying generic characteristic. In this paper, we searched for  universal feedback of 
	non-Markovian noise on quantum correlations (QC) of random two-qubit states. In particular,  a single qubit  (single-sided) or both the qubits  (double-sided) of 
	Haar uniformly simulated states with different ranks  are subjected to two different non-Markovian 
	channels, dephasing and  depolarizing channels  with varying non-Markovian strength. And we choose logarithmic negativity and quantum discord as  QC measures
	for investigations. We found that  broad common features regarding the response  of noise and non-Markovianity on QCs of random states emerge 
	although there are some differences in the observations of two different channels as well as in single- and 
	double-sided channels. To capture these aspects, we introduced certain quantities which can help us to visualize  the response to noise 
	and non-Markovianity of QCs of random states.  
	
	In the case of dephasing channel, we observed that  if a single  qubit is affected by noise,  entanglement is, in general,  reviving more number of times  than 
	the situation when both the qubits are subjected to noisy channels. Similar behaviour has been observed for QD. In a similar spirit, when  both the qubits are sent through depolarizing channels, entanglement regenerates after collapses while 
	no revival is observed for a single-sided case. In both  situations, we observe that  more noisy scenario leads to more revival of QCs -- 
	this is possibly due to the presence of non-Markovianity in the channels, which we call as constructive feedback of non-Markovian noise. The results are supported both by analytical and numerical means. The fact that the mean increases with the increase of non-Markovian power justifies the constructive effect.
	Moreover, we noticed that depolarizing channel leads to a  more revival of quantum discord than that of the dephasing ones, in double-sided channels.  It does not hold for entanglement -- especially, for a single-sided depolarizing channel, entanglement of random states
	does not become nonvanishing after collapse which can be argued intuitively.  In near future, it will be interesting to 
	derive a complementary relation between the noise parameter and the strength of non-Markovianity for a given channel and for a fixed QCs of the initial state. Such a study can help us to compare  different channels quantitatively. 
	
	On the other hand, we found that both for entanglement and quantum discord, the average value of threshold noise at which QCs of the output states
	collapse for the first time decreases with the increase of non-Markovian parameter of the channels. This observation is independent of the channels
	considered and the rank of the random input states. We noticed that similar picture emerges also for the mean critical noise strength at which 
	QCs revive after the first collapse.  It implies that although non-Markovianity does not give any advantage to preserve QCs initially, it surely induces the first regeneration of QCs, on average. In case of QD, the difference between mean critical noise for collapse and regeneration is very less in comparison to that for entanglement. It implies that QD revives easily after collapse whereas it is hard  for entanglement  to revive even through non-Markovianity which can  be intuitively  understood from the natures of QC measures.
	Moreover, we reported that a certain amount of average initial QCs in random states along with a moderate values of non-Markovian noise is  responsible for regeneration of quantum correlations. We believe that our investigations 
	shed light on how the resources are affected due to the competition between the damping parameter and the non-Markovianity in the noisy channels.
	

	\section{Acknowledgement}
	We acknowledge the support from Interdisciplinary Cyber Physical Systems
	(ICPS) programme of the Department of Science and Technology (DST), India,
	Grant No.: DST/ICPS/QuST/Theme- 1/2019/23 (General Project No. Q-121).
	The authors acknowledge Tamoghna Das, for his help regarding the various plots. S. Gupta thanks the hospitality of Harish-Chandra Research Institute. Numerical results have been obtained using the Quantum Information and Computation library (QIClib) (https://titaschanda. github.io/QIClib)  and computational work for this study was carried out at the cluster computing facility in the Harish-Chandra Research Institute (http://www.hri.res.in/cluster).
	

	\section*{Appendix: Quantum correlation Measures}
	
	We here work with two kinds of quantum correlation measures, namely logarithmic negativity, an entanglement measure and quantum discord, a QC measure which has its origin different from entanglement. 
	
	\emph{Logarithmic Negativity. }
	Based on partial transposition criteria \cite{PT}, logarithmic negativity  (LN)  \cite{Vidal2002}, a computable entanglement measure,  of a given bipartite state $\rho_{AB}$ is defined as
	\begin{equation}
		E (\rho_{AB})=\log_2||\rho_{AB}^{T_A}||=\log_2(2N (\rho_{AB}) +1)
	\end{equation}
	where $\rho_{AB}^{T_A}$ is the state after partial transposition with respect to subsystem   A, \( || . ||\) denotes the trace norm, while \( N (\rho_{AB})\) is called the negativity which is the  sum of absolute value of  negative eigenvalues of $\rho_{AB}^{T_A}$. Notice that \(N(\rho_{AB})\) can also be a valid of measure of entanglement which we will use to derive analytical results.  

	\emph{Quantum discord.} It  quantifies quantum correlation present in \(\rho_{AB}\) which is independent from entanglement \cite{QD, discordreviewModi, discordreviewBera} and has originated from the concept of classical information theory \cite{CoverThomasbook}.  Classically, mutual information is defined as
	\begin{equation}
		I(X:Y) = H(X) + H(Y) - H(X,Y) = H(X) - H(X|Y)
	\end{equation}
	where $X$ is the random variable having probability distribution, $\{p_x\}$, $H(X) = - \sum_x p_x \log_2 (p_x)$ is the Shannon entropy and similarly $H(Y)$.  \(H(X,Y)\) is the Shannon entropy of the joint probability distribution of $X$ and $Y$ and  $H(X|Y) =  H(X, Y) - H(Y)$ is the conditional entropy.
	Switching to the quantum realm, mutual information of \(\rho_{AB}\) reads as
	\begin{equation}
		I (\rho_{AB}) = S(\rho_A) + S(\rho_B) - S(\rho_{AB})
	\end{equation}
	where $S(\rho) = - \mbox{tr} (\rho \log_2 \rho)$ is the Von-Neumann entropy, and \(\rho_i, \, i = A, B\) are the reduced density matrices of \(\rho_{AB}\).
	If the second definition of mutual information, that involving conditional entropy, is recast in terms of the Von-Neumann entropy, we get
	\begin{equation}
		J(\rho_{AB}) = S(\rho_A) - S(\rho_{A|B})
	\end{equation}
	which can both be positive as well as negative.
	Therefore, the second term in the above equation can be modified as 
	\begin{equation}
		S(\rho_{A|B}) = \min_{\{\Pi^B_k\} } \sum_k p_k S(\rho_{A|k}) 
	\end{equation}
	where the minimization is taken over all projective rank-1 measurements, \(\{\Pi^B_k\}\), on the subsystem B. The post-measurement state is $\rho_{A|k}$, which is obtained with probability $p_k$, both of which, can be expressed as
	\begin{eqnarray}
		\rho_{A|k} = \mbox{tr}_B (\Pi^B_k \rho_{AB} \Pi^B_k)/p_k, \, 
		p_k = \mbox{tr}(\Pi^B_k \rho_{AB} \Pi^B_k).
	\end{eqnarray}
	The difference between  quantum mutual information and  \(J(\rho_{AB})\) leads to the definition of  quantum correlation measure called quantum discord, given by
	\begin{equation}
		D(\rho_{AB}) = I(\rho_{AB}) - \max_{\{\Pi^B_k\}} J(\rho_{AB}),
	\end{equation}
	where the first and the second terms respectively can be interpreted as total and classical correlations. 


\begin{thebibliography}{1}
\bibitem{randompure} The parameters chosen for two random pure states are 
		\begin{eqnarray}
			\{a_i, b_i\}_{i=1}^{4} \in &&\{(0.393254, 1.81283), 0.931627, 1.43439),  \nonumber \\
			&& (-0.440117, -1.16197), (1.01733, -1.74792)\} \nonumber
			\mbox{and}\nonumber \\
			&&\{(-1.52442, -0.4851), (-1.01821, -0.631922), \nonumber \\
			&& (-0.135748, 0.391063), (1.31956, 1.28777)\}.\nonumber
		\end{eqnarray}
\bibitem{randompuremix} The parameters chosen for  random pure states are 
		\begin{eqnarray}
			\{(a_{ij}, b_{ij})\}_{i,j=0,1} \in && \{(-1.57089, -1.61154), (-0.535571, -0.523942), \nonumber \\
			&&  (0.621694, 1.09599), (-0.585193,  -1.09819)\} \nonumber
		\end{eqnarray}
		while for rank-2 states, eight pairs of \((a_{ijk}, b_{ijk})\) are chosen as follows:
		\begin{eqnarray}
			&&\{(a_{ijk}, b_{ijk})\}_{i,j,=0,1} \in  \{(0.817452, 3.00601), (-0.316579, 0.79375),\nonumber \\
			&& (-0.713834, 0.30082), (0.157244, 0.675026), (0.885021, 0.503886), \nonumber \\
			&& (1.01862, -0.642617), (-1.73205, -1.12481), (-0.11353, -2.00412)\}.\nonumber 
		\end{eqnarray}
\bibitem{normdistri} Normalized frequency distribution of  QC measure, \(\mathcal{Q}\) can be obtained by computing \(\frac{\mathcal{N}(\mathcal{Q})}{\mathcal{N}}\) where \(\mathcal{N}(\mathcal{Q})\) is the number of random states with the amount of QCs, \(\mathcal{Q}\) and \(\mathcal{N}\) is the total number of states generated Haar uniformly.
\end{thebibliography}
\end{document}